\documentclass[pre,twocolumn,amsmath,amssymb,nofootinbib,floatfix,superscriptaddress]{revtex4}

\usepackage{graphicx,bm,xcolor, bbold}
\usepackage[normalem]{ulem}
\usepackage{hyperref}

\makeatletter
\def\graphicscale{\twocolumn@sw{0.3}{0.4}}
\def\graphicthreescale{\twocolumn@sw{0.3}{0.4}}

\begin{document}

\title{Quantum critical behaviors and decoherence of weakly coupled
  \\ quantum Ising models within an isolated global system}

\author{Alessio Franchi}
\affiliation{Dipartimento di Fisica dell'Universit\`a di Pisa
        and INFN, Largo Pontecorvo 3, I-56127 Pisa, Italy}

\author{Andrea Pelissetto}
\affiliation{Dipartimento di Fisica dell'Universit\`a di Roma Sapienza
        and INFN Sezione di Roma I, I-00185 Roma, Italy}

\author{Ettore Vicari} 
\affiliation{Dipartimento di Fisica dell'Universit\`a di Pisa
        and INFN, Largo Pontecorvo 3, I-56127 Pisa, Italy}

\date{\today}

\begin{abstract}
We discuss the quantum dynamics of an isolated composite system
consisting of weakly interacting many-body subsystems.  We focus on
one of the subsystems, ${\cal S}$, and study the dependence of its
quantum correlations and decoherence rate on the state of the
weakly-coupled complementary part ${\cal E}$, which represents the
{\em environment}.  As a theoretical laboratory, we consider a
composite system made of two stacked quantum Ising chains, locally and
homogeneously weakly coupled. One of the chains is identified with the
subsystem ${\cal S}$ under scrutiny, and the other one with the {\em
  environment} ${\cal E}$.  We investigate the behavior of ${\cal S}$
at equilibrium, when the global system is in its ground state, and
under out-of-equilibrium conditions, when the global system evolves
unitarily after a soft quench of the coupling between ${\cal S}$ and
${\cal E}$.  When ${\cal S}$ develops quantum critical correlations in
the weak-coupling regime, the associated scaling behavior crucially
depends on the quantum state of ${\cal E}$, whether it is
characterized by short-range correlations (analogous to those
characterizing disordered phases in closed systems), algebraically
decaying correlations (typical of critical systems), or long-range
correlations (typical of magnetized ordered phases).  In particular,
different scaling behaviors, depending on the state of $\cal E$, are
observed for the decoherence of the subsystem ${\cal S}$, as
demonstrated by the different power-law divergences of the decoherence
susceptibility that quantifies the sensitivity of the coherence to the
interaction with~${\cal E}$.
\end{abstract}

\maketitle


\section{Introduction}
\label{intro}

The recent progress on the nano-scale control of physical systems has
opened the road to investigations of the quantum properties and of the
coherent quantum dynamics of coupled systems, addressing also issues
concerning the relative decoherence and the energy flow among the
various subsystems~\cite{Zurek-03}.  These investigations improve our
understanding of the emergence of interference and entanglement, which
is useful for quantum-information purposes~\cite{NielsenChuang}, or
for enhancing the efficiency of energy conversion in complex
networks~\cite{Lambert-13}.  The presence of different quantum phases
and the development of critical behaviors in interacting subsystems
are expected to play a crucial role for the emergence of new phenomena
in the equilibrium and out-of-equilibrium dynamics of isolated and
open quantum systems~\cite{Sachdev-book,RV-21}.

If we consider a quantum system made up of various components, any
subsystem can be seen as an effective bath for the other ones.  In
this context, one may study the quantum dynamics of an open system
subject to the interaction with the environment, while the global
system (composed of the open system and its environment) evolves
unitarily.  These issues have been already addressed within some
paradigmatic, relatively simple, composite models, such as the
so-called {\em central-spin} models, where one or few qubits are
globally coupled to an environmental many-body system~\cite{Zurek-82,
  CPZ-05, QSLZS-06, RCGMF-07, CFP-07, CP-08, Zurek-09, DQZ-11, NDD-12,
  SND-16, V-18, FCV-19, RV-19, RV-21}, and {\em sunburst} models where
sets of isolated qubits are locally coupled to a many-body
system~\cite{FRV-22,FRV-22-2}.  The decoherence properties of the
subsystems crucially depend on the large-scale features of the state
they are in, for instance, on whether the subsystem is in an ordered
or a disordered quantum phase, or it is close to a critical point,
where large-scale critical correlations develop~\cite{Sachdev-book}.

In this paper we focus on the {\em open} dynamics of one many-body
subsystem ${\cal S}$ weakly coupled with a complementary {\em
  environment} ${\cal E}$.  We study the dependence of the critical
behavior of ${\cal S}$ on the coupling between ${\cal S}$ and ${\cal
  E}$, and on the state of the {\em environment} ${\cal E}$, which can
be controlled by varying the Hamiltonian parameters.

\begin{figure}
  \includegraphics[width=0.95\columnwidth]{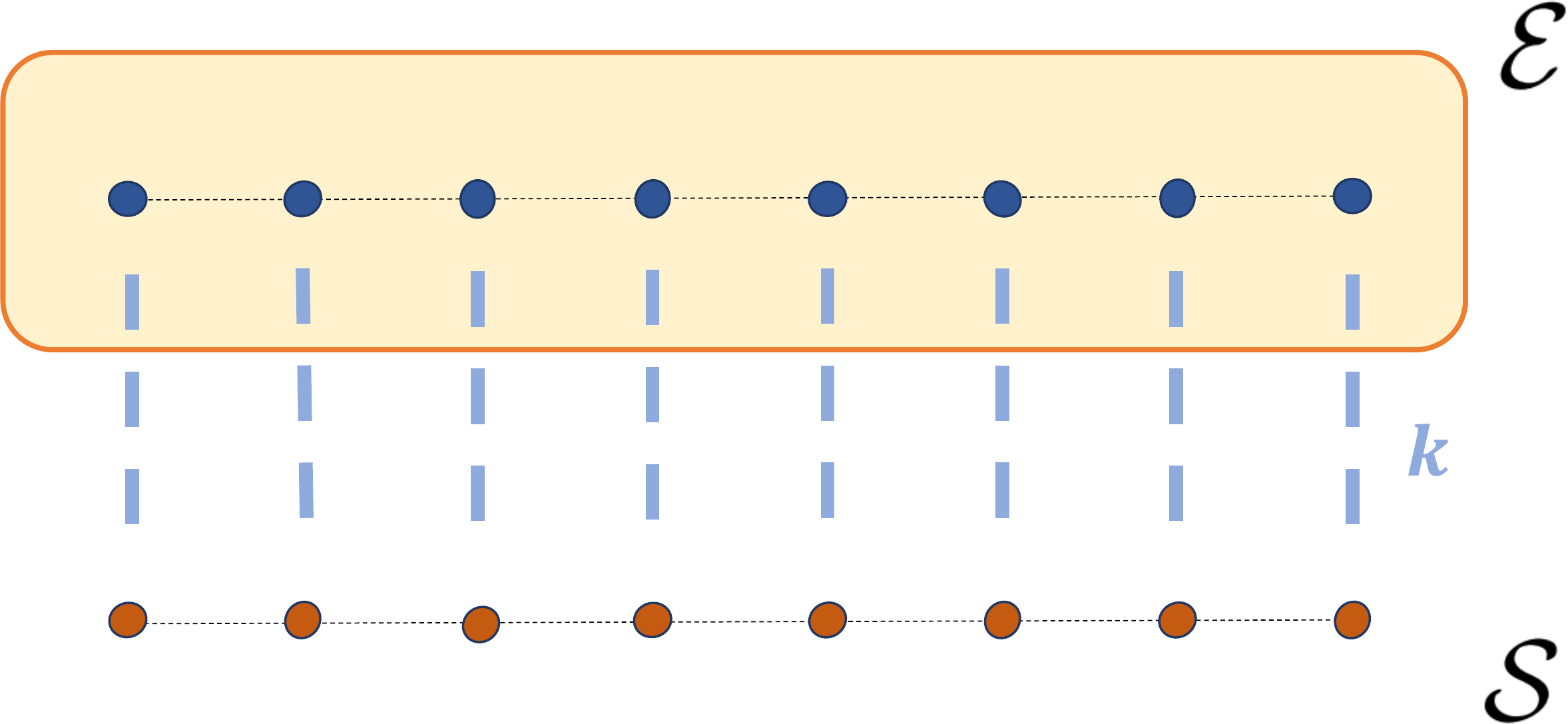}
  \caption{Sketch of a system made of two stacked Ising
    chains, weakly coupled by local and homogenous interactions
    controlled by a parameter $\kappa$.  One of the chains represents
    the subsystem ${\cal S}$, while the other is the
    {\em environment} ${\cal E}$.  }
  \label{sketchsystem}
\end{figure}

As a theoretical laboratory, we consider two stacked one-dimensional
Ising chains, locally and homogeneously weakly coupled, as sketched in
Fig.~\ref{sketchsystem}.  One of the chains represents the subsystem
${\cal S}$, the other one is the {\em environment} ${\cal E}$. The
Hamiltonian parameters of the two chains (for example, the Hamiltonian
coefficient of the external transverse field) differ, so that ${\cal
  S}$ and ${\cal E}$ may be in different quantum phases.  We discuss
how a weak interaction between ${\cal S}$ and ${\cal E}$ (controlled
by one parameter $\kappa$ that vanishes when ${\cal S}$ and ${\cal E}$
are decoupled) affects the quantum scaling behaviors and the
decoherence rate of ${\cal S}$, close to critical transitions.  We
discuss static properties, assuming that the global system is in its
ground state, and the out-of-equilibrium behavior when a slow quench
of the interactions between ${\cal S}$ and ${\cal E}$ is performed.
We show that the scaling behavior in the weak-coupling regime depends
on the quantum state of ${\cal E}$. More precisely, a disordered,
critical, ordered environment, characterized by short-range,
algebraically decaying, and long-range correlations, respectively,
differently affects the critical scaling behavior of $\cal S$.

To characterize the scaling behavior in the presence of a coupling
between ${\cal S}$ and ${\cal E}$ controlled by a parameter $\kappa$,
we use the renormalization-group (RG) approach, which provides the
natural theoretical framework to effectively describe the behavior of
systems in proximity of quantum transitions, see, e.g.,
Refs.~\cite{Sachdev-book,RV-21}. We focus, in particular, on the
large-size behavior of the system, deriving finite-size scaling (FSS)
relations, which are largely independent of the microscopic
details. Therefore, they hold in widely different systems and in very
different physical contexts.  Moreover, they allow us to describe
complex phenomena using a relatively small number of relevant
variables, providing a notable simplification of the analysis.

As we mentioned above, in the presence of a coupling between $\cal S$
and $\cal E$, some features of the critical behavior of $\cal S$
depend on whether $\cal E$ is disordered, critical, or ordered.  For
instance, the sensitivity of the coherence properties of $\cal S$ to
the coupling strength $\kappa$ is significantly different in these
three cases.  Such a sensitivity can be effectively quantified by
using the the susceptibility $\chi_Q$ of the decoherence factor of
${\cal S}$ with respect to the parameter $\kappa$, since this quantity
shows a different power-law divergence with the size $L$ of the system
in the three cases mentioned above. The RG predictions derived in this
work have been confirmed by the FSS analysis of numerical results for
stacked Ising chains.

In a dynamic perspective, the study of the phase diagram and of the
scaling properties of the global system determines the adiabatic limit
of a slow dynamics for a finite-size system (we recall that
finite-size many-body systems are generally gapped).  We also extend
the discussion to out-of-equilibrium conditions, determining the
effects of an instantaneous quench of the coupling between ${\cal S}$
and ${\cal E}$. Again we use a RG framework, deriving general dynamic
FSS relations that extend those obtained for the system in
equilibrium.

The paper is organized as follows. In Sec.~\ref{models} we define the
system we consider, composed of two coupled (stacked) $d$-dimensional
quantum Ising systems. In Sec.~\ref{obs} we introduce the observables
we use to characterize the critical properties of the subsystem ${\cal
  S}$.  In Sec.~\ref{smallqsca} we derive general FSS relations that
characterize the equilibrium behavior of the subsystem ${\cal S}$ in
the presence of a weak coupling with the environment ${\cal E}$.  The
different RG scaling ans\"atze, which depend on the quantum phase of
the environment, are supported by numerical results for coupled
quantum Ising chains. In Sec.~\ref{phdia} we discuss the general
features of the phase diagram for finite values of the coupling
$\kappa$ between ${\cal S}$ and ${\cal E}$.  In Sec.~\ref{dynsca} we
extend the discussion to out-of-equilibrium dynamic processes,
considering a soft quench of the interaction between ${\cal S}$ and
${\cal E}$.  Finally, in Sec.~\ref{conclu} we summarize and draw our
conclusions. App.~\ref{AppA} provides a mean-field analysis of the
phase diagram of stacked Ising systems.  App.~\ref{AppB} reports exact
results in some limiting cases.

\section{Coupled quantum Ising systems}
\label{models}

We consider a system composed of two interacting (stacked)
$d$-dimensional quantum Ising models: one of them is identified as the
subsystem ${\cal S}$ under observation and the other one as the {\em
  environment} ${\cal E}$. The Hamiltonian of the global system is
\begin{eqnarray}
  H = H_{\cal S}(J,g) + H_{\cal E}(J_e,g_e) + H_{\cal SE}(\kappa) \,,
  \label{twosys}
\end{eqnarray}
where
\begin{eqnarray}
  &&H_{\cal S}(J,g) = - J\sum_{\langle {\bm x}{\bm y}\rangle}
  \sigma^{(1)}_{{\bm x}} \sigma^{(1)}_{{\bm y}} - g \sum_{\bm x}
  \sigma^{(3)}_{{\bm x}}\,, \label{HSdef}\\
  &&H_{\cal E}(J_e,g_e)
  = - J_e \sum_{\langle {\bm x}{\bm y}\rangle} \tau^{(1)}_{{\bm
      x}} \tau^{(1)}_{{\bm y}} - g_e \sum_{\bm x}
  \tau^{(3)}_{{\bm x}}\,, \label{HEdef}\\
  &&H_{\cal SE}(\kappa) = -
  \kappa \, \sum_{\bm x} \sigma^{(1)}_{{\bm x}}
  \tau^{(1)}_{{\bm x}}\,,\label{HSEdef}
\end{eqnarray}
where ${\bm x}$ are the sites of a cubic-like lattice of size $L^d$,
${\langle {\bm x} {\bm y} \rangle}$ indicates nearest-neighbor sites,
$\sigma^{(k)}_{\bm x}$ and $\tau^{(k)}_{\bm x}$ are two independent
sets of Pauli matrices.  In the following we consider generic boundary
conditions, for example open or periodic boundary conditions (OBC and
PBC, respectively).  The coupling $\kappa$ controls the strength of
the interactions between the subsystems ${\cal S}$ and ${\cal E}$,
while the Hamiltonian parameters $J_e$ and $g_e$ allow us to control
the quantum state of the environment (in the regime of weak coupling
between ${\cal S}$ and ${\cal E}$). To reduce the number of input
parameters, we set
\begin{equation}
J=J_e=1\,,
\label{jje}
\end{equation}
which does not limit the generality of our discussion (unless one is
interested in some particular limits that we do not consider).  For
$d=1$ we obtain the stacked Ising chains sketched in
Fig.~\ref{sketchsystem}.

For $\kappa = 0$, the system is invariant under the 
${\mathbb Z}_2\otimes {\mathbb Z}_2$ group of transformations 
that independently change the signs of
the $\sigma^{(1)}_{\bm x}$ and $\tau_{\bm x}^{(1)}$ operators.
The interaction Hamiltonian $H_{\cal SE}$ breaks this invariance, 
leaving only a global
${\mathbb Z}_2$ symmetry under the simultaneous transformations
\begin{eqnarray}
  &\sigma^{(1/2)}_{\bm x} \to -\sigma^{(1/2)}_{\bm x} \,, \qquad
&\sigma^{(3)}_{\bm x} \to \sigma^{(3)}_{\bm x} \,,
\label{globalZ2}\\
&\tau^{(1/2)}_{\bm x} \to -\tau^{(1/2)}_{\bm x} \,, \qquad
&\tau^{(3)}_{\bm x} \to \tau^{(3)}_{\bm x} \,.
\nonumber
\end{eqnarray}
Note that, if we only change the sign of one the longitudinal spin
operators, i.e., $\sigma^{(1)}_{\bm x}\to - \sigma^{(1)}_{\bm x}$ or
$\tau^{(1)}_{\bm x}\to -\tau^{(1)}_{\bm x}$, we obtain the same
Hamiltonian with $\kappa$ replaced by $-\kappa$. Thus, the phase
diagram does not depend on the sign of $\kappa$. Without loss of
generality, we assume $\kappa \ge 0$.  Moreover, at fixed $\kappa$,
the phase diagram of the global system is invariant under the exchange
$g\leftrightarrow g_e$, due to the fact that the two subsystems ${\cal
  S}$ and ${\cal E}$ are identical apart from their transverse-field
parameters $g$ and $g_e$.

A simplified model is obtained for $g=g_e$.  In this case, the global
system has an additional ${\mathbb Z}_2$ symmetry, as it is invariant
under the interchange $\sigma_{\bm x} \leftrightarrow \tau_{\bm x}$.
Thus, the symmetry group enlarges to ${\mathbb Z}_2 \otimes {\mathbb
  Z}_2 \otimes {\mathbb Z}_2$ for $\kappa = 0$ and to ${\mathbb Z}_2
\otimes {\mathbb Z}_2$ when the two systems are coupled
($\kappa\not=0$).

It is worth noting that the symmetry properties of the global system
change, if the coupling Hamiltonian involves the transverse spin
operators, i.e., if $H_{\cal SE}$, defined in Eq.~(\ref{HSEdef}), is
replaced by
\begin{equation}
\widetilde{H}_{\cal SE}=-\widetilde\kappa
\sum_{\bm x} \sigma^{(3)}_{{\bm x}} \tau^{(3)}_{{\bm x}}\,,
\label{altint}
\end{equation}
where $\widetilde\kappa$ is the parameter controlling the strength of
the interaction.  Indeed, such a coupling term preserves the ${\mathbb
  Z}_2\otimes {\mathbb Z}_2$ symmetry present for $\widetilde\kappa =
0$. Note that, in the symmetric case $g=g_e$, the global model with
the interacting term $\widetilde{H}_{\cal SE}$ is equivalent to the
so-called quantum Ashkin-Teller model, see, e.g.,
Refs.~\cite{Fradkin-84,Shankar-85}. In this paper we only consider
models coupled through their longitudinal spin operators, such as in
Eq.~(\ref{HSEdef}), breaking the independent ${\mathbb Z}_2$
invariance of the two subsystems. The alternative case, corresponding
to the coupling $\widetilde{H}_{\cal SE}$, is also worth
investigating, as it would provide insights on the quantum dynamics,
when the interaction term between ${\cal S}$ and ${\cal E}$ does not
break any symmetry of the isolated system.  We do not pursue it in
this paper.

When the interaction between the subsystems vanishes, i.e., when
$\kappa=0$, one recovers two decoupled $d$-dimensional quantum Ising
systems. Therefore, it is useful to recall that $d$-dimensional
quantum Ising systems, described by the Hamiltonian (\ref{HSdef})
supplemented by an additional longitudinal term $H_h=-h \sum_{\bm x}
\sigma_{\bm x}^{(1)}$, undergo a quantum continuous transition at a
finite value $g=g_{\cal I}$ and $h=0$; see, e.g.,
Ref.~\cite{Sachdev-book,RV-21}.  The corresponding quantum critical
behavior belongs to the $(d+1)$-dimensional Ising universality class.
In particular, we have $g_{\cal I}=1$ for the one-dimensional Ising
chain.  The relevant parameters $r\equiv g-g_{\cal I}$ and $h$,
associated with the transverse and longitudinal spin operators
$\sigma_{\bm x}^{(3)}$ and $\sigma_{\bm x}^{(1)}$, represent the
leading even and odd RG perturbations at the $(d+1)$-dimensional Ising
fixed point.  Their RG dimensions are $y_r=1/\nu$ and
$y_h$, respectively,
so that the length scale $\xi$ of the critical modes behaves as
$\xi\sim |g-g_{\cal I}|^{-\nu}$ for $h=0$, and $\xi\sim |h|^{-1/y_h}$
for  $g=g_{\cal I}$.  The RG exponents are exactly known for
one-dimensional systems: $y_r=1$ and $y_h = 15/8$, see, e.g.,
Ref.~\cite{Sachdev-book}.  Accurate estimates are available for
two-dimensional quantum Ising systems, see, e.g.,
Refs.~\cite{PV-02,GZ-98,CPRV-02,Hasenbusch-10,KPSV-16,KP-17,Hasenbusch-21};
Ref.~\cite{KPSV-16} reports $y_r= 1.58737(1)$ and $y_h = 2.481852(1)$.
For $d=3$, the critical exponents take their mean-field values,
$y_r=2$ and $y_h=3$; moreover, the critical singular behavior is
characterized by additional multiplicative logarithmic
factors~\cite{Sachdev-book,RV-21,PV-02}.  The dynamic exponent $z$,
controlling the vanishing of the gap $\Delta\sim \xi^{-z}$ at the
transition point, is 1 in any dimension. For later use, we recall that
the RG dimension $y_\phi$ of the longitudinal spin operator
$\sigma_x^{(1)}$ (it represents the order parameter of the model) is
given by $y_\phi=d + z - y_h=(d+z-2+\eta)/2$, where $\eta$ is the
critical exponent characterizing the spatial decay of the critical
correlations. Using the above-reported results for $y_h$, we have
$y_\phi=1/8$ for $d=1$, $y_\phi=0.518148(1)$ for $d=2$, and $y_\phi=1$
for $d=3$.

\section{Observables at equilibrium}
\label{obs}

To study the {\em equilibrium} properties of the subsystem ${\cal S}$
when the global system is in the ground state $|\Psi_0\rangle$,
considering ${\cal E}$ as the environment, we introduce the reduced
density matrix of ${\cal S}$
\begin{equation}
  \rho_{\cal S} = {\rm Tr}_{\cal E} \, \big[|\Psi_0\rangle \langle
    \Psi_0|\big]\,,
  \label{rhoIA}
\end{equation}
where ${\rm Tr}_{\cal E} [\,\cdot\,]$ denotes the partial trace over
the Hilbert space associated with the subsystem ${\cal E}$.

The coherence properties of ${\cal S}$ can be quantified through the
purity $P$, the corresponding R{\'e}nyi entanglement entropy $S$, and
the decoherence factor $Q$, defined as
\begin{equation}
  P = {\rm Tr} \big[ \rho_{\cal S}^2 \big] \,,
  \qquad S = - {\rm ln} \,
  P \,,\qquad Q = 1 - P \,.
  \label{purentdec}
\end{equation}
Exploiting the Schmidt decomposition for bipartitions of pure states,
one can easily prove that the purity $P$ of the subsystem ${\cal S}$
equals that of the complementary environment ${\cal E}$.  The
decoherence factor varies between zero (corresponding to $P=1$ and
$S=0$, for a pure reduced state) and 1 (corresponding to $P = 0$, for
a completely incoherent many-body state).

To quantify the loss of coherence of the subsystem ${\cal S}$ due to a
weak interaction term $H_{\cal SE}$, we look at the behavior of $Q$
for small values of $\kappa$.  Since $Q$ is an even function of
$\kappa$, assuming analyticity at $\kappa=0$ (which is certainly true
for finite-size systems), we can expand it as
\begin{equation}
  Q = {1\over 2} \kappa^2 \chi_Q + O(\kappa^4)\,, \qquad
  \chi_Q\equiv {\partial^2 Q \over \partial \kappa^2}\bigg|_{\kappa=0}\,,
\label{chiqdef}
\end{equation}
where $\chi_Q$ represents the decoherence susceptibility with respect
to the coupling $\kappa$.

We also consider the correlations of the spin operators
$\sigma_x^{(1)}$.  Due to the global ${\mathbb Z}_2$ symmetry, we have
\begin{equation}
{\rm Tr} \big[\rho_{\cal S}\, \sigma_{\bm x}^{(1)}
  \big] =0\,.\label{vanishM}
\end{equation}
The two-point correlation function can be written as
\begin{equation}
  G({\bm x},{\bm y}) \equiv {\rm Tr} \big[ \rho_{\cal S} \,\sigma_{\bm x}^{(1)}
    \sigma_{\bm y}^{(1)} \big] \,.\label{gxy}
\end{equation}
We consider odd values of $L$, we set $L=2\ell+1$, and choose
coordinates such that $-\ell \le x_i \le \ell$, so that we can
identify a central site ${\bm x}_0$ with vanishing coordinates (this
coordinate system is particularly convenient in the case of OBC).  We
consider a susceptibility and second-moment correlation length,
defined as
\begin{equation}
\chi = \sum_{\bm x}   G({\bm x}_0,{\bm x})\,,\quad 
\xi^2 = {1\over 2 d \chi}\sum_{\bm x} {\bm x}^2
G({\bm x}_0,{\bm x})\,.
\label{chixidef}
\end{equation}
The ratio 
\begin{equation}
  R_\xi = \xi/L
  \label{rxidef}
\end{equation}
is a RG invariant quantity. In the FSS limit, it scales as 
$R_\xi(g,L) \approx {\cal R}(rL^{y_r})$,
where $r=g-g_{\cal I}$ and ${\cal R}$ is a function that is universal
apart from a rescaling of its argument~\cite{PV-02,CPV-14}.  Moreover,
the susceptibility $\chi$ behaves as 
$\chi \approx \,L^{d-2 y_\phi} {\cal C}(r L^{y_r})$,
where $d-2 y_\phi=2-z-\eta=1-\eta$, 
or, equivalently, as 
\begin{equation}
  \chi \approx L^{1-\eta} a_\chi F_\chi(R_\xi)\,.
  \label{chiscaisi}
  \end{equation}
The function $F_\chi(R_\xi)$ is universal, while $a_\chi$ is a nonuniversal
constant that depends on the model parameters.

\section{Scaling behaviors for weakly coupled subsystems}
\label{smallqsca}

We now discuss how the coupling term $H_{\cal SE}$ affects the quantum
critical properties of ${\cal S}$ for small values of $\kappa$.  Using
RG arguments, we show that ${\cal S}$ develops different scaling
behaviors, that depend on the state of the environment ${\cal E}$,
controlled by the Hamiltonian parameter $g_e$, cf. Eq.~(\ref{HEdef}).
We refer here to the state of $\cal E$ for $\kappa = 0$. Indeed, the
addition of a coupling term $H_{\cal SE}$ also changes the
properties of the environment.  We distinguish three cases: (i) the
environment ${\cal E}$ is disordered, $g_e>g_{\cal I}$; (ii) ${\cal
  E}$ is in the critical regime, $g_e\approx g_{\cal I}$; (iii) the
environment is ordered (magnetized), i.e., $g_e<g_{\cal I}$.

The predicted scaling behaviors will be compared with the results of
numerical FSS analyses for the one-dimensional model (\ref{twosys}),
i.e., for two stacked Ising chains, sketched in
Fig.~\ref{sketchsystem}. To compute correlation functions, we use the
density-matrix RG (DMRG) algorithm with OBC, which allows us to obtain
results for systems of size up to $L\approx 40$.  As for the
implementation, we use matrix-product-state (MPS) algorithms taken
from the iTensor library~\cite{FWS-20}.  The DMRG algorithm is very
convenient to compute coherence properties for left-right bipartitions
of the system. In principle, it could also be used to compute $Q$ in
our case, by considering $\cal S$ and $\cal E$ as the left and right
ordering of the ladder model that we consider. However, in this type
of implementation, one would generate nonlocal interactions between
the two subsystems, making the algorithm inefficient.  Therefore, the
decoherence factor $Q$ has been computed by performing an exact
diagonalization of the global Hamiltonian. Of course, smaller systems
can be considered (we obtained results up $L\approx 10$). In this case
we used PBC.

\subsection{Disordered environment}
\label{disenv}

Let us first assume that the environment ${\cal E}$ is disordered for
$\kappa=0$.  It presents only short-range correlations, so that it may
be effectively considered as a collection of a large number of
independent subsystems. For an Ising system, such as the one described
by the Hamiltonian (\ref{HEdef}), the environment ${\cal E}$ is
disordered for $g_e>g_{\cal I}$.  We focus on the subsystem ${\cal
  S}$, and, in particular, on the scaling behavior of its decoherence
properties for small values of the coupling $\kappa$.

\subsubsection{Scaling behavior for small $\kappa$}
\label{disordsca}

For $g_e>g_{\cal I}$, we conjecture that the interaction $H_{\cal SE}$
between ${\cal S}$ and ${\cal E}$ is an irrelevant perturbation at the
quantum critical point of the subsystem ${\cal S}$.  Under this
hypothesis, the subsystem ${\cal S}$ has an Ising critical transition
also for finite $\kappa$, at a critical point $g_c(\kappa)$ that
depends on $\kappa$.  Since the coupling $\kappa$ is irrelevant, there
is only one relevant operator also for $\kappa\not=0$.  We indicate
the corresponding scaling field with $u_r$ ($u_r\sim r$ in the absence
of coupling $\kappa$). Its RG dimension $y_r$ is the same as that of
the thermal operator in the Ising universality class, see at the end
of Sec.~\ref{models}.

We recall that the scaling fields associated with the RG perturbations
are analytic functions of the model
parameters~\cite{Wegner-76,PV-02,CPV-14,RV-21}. In the case at hand,
in which there is only one relevant RG perturbation, the singular part
of the free-energy density in the zero-temperature and FSS limit is
expected to scale as~\cite{CPV-14,RV-21}
\begin{eqnarray}
F_{\rm sing}(g,g_e>g_{\cal I},\kappa,L) \approx L^{-(d+z)}{\cal F}(W_s) \,,
  \label{freeendis}
\end{eqnarray}
where
\begin{eqnarray}
W_s = u_r L^{y_r}\,.\label{Wsdefdis}
\end{eqnarray}
Correspondingly, any RG invariant quantity $R$, such as $R_\xi$
defined in Eq.~(\ref{rxidef}) or the decoherence factor $Q$ defined in
Eq.~(\ref{purentdec}), is expected to asymptotically scale
as~\cite{CPV-14}
\begin{equation}
  R \approx F_R(W_s)\,.
  \label{Rscal}
\end{equation}
The scaling field $u_r$ is an analytic function of $g$, $\kappa$, and
$g_e$, such that $u_r\sim r$ for $\kappa=0$. Due to the symmetry of
the phase diagram under $\kappa\to -\kappa$, for small values of $r$
and $\kappa$ it behaves as
\begin{equation}
  u_r(r,\kappa) \approx r + b\, \kappa^2\,,\qquad r = g - g_{\cal
    I}\,,
  \label{ug}
\end{equation}
where the nonuniversal constant $b$ depends on the environment
coupling $g_e$, and we fixed an arbitrary normalization requiring
$u_r\approx r$ for $\kappa = 0$.

\begin{figure}
    \centering
    \includegraphics[width=0.95\columnwidth]{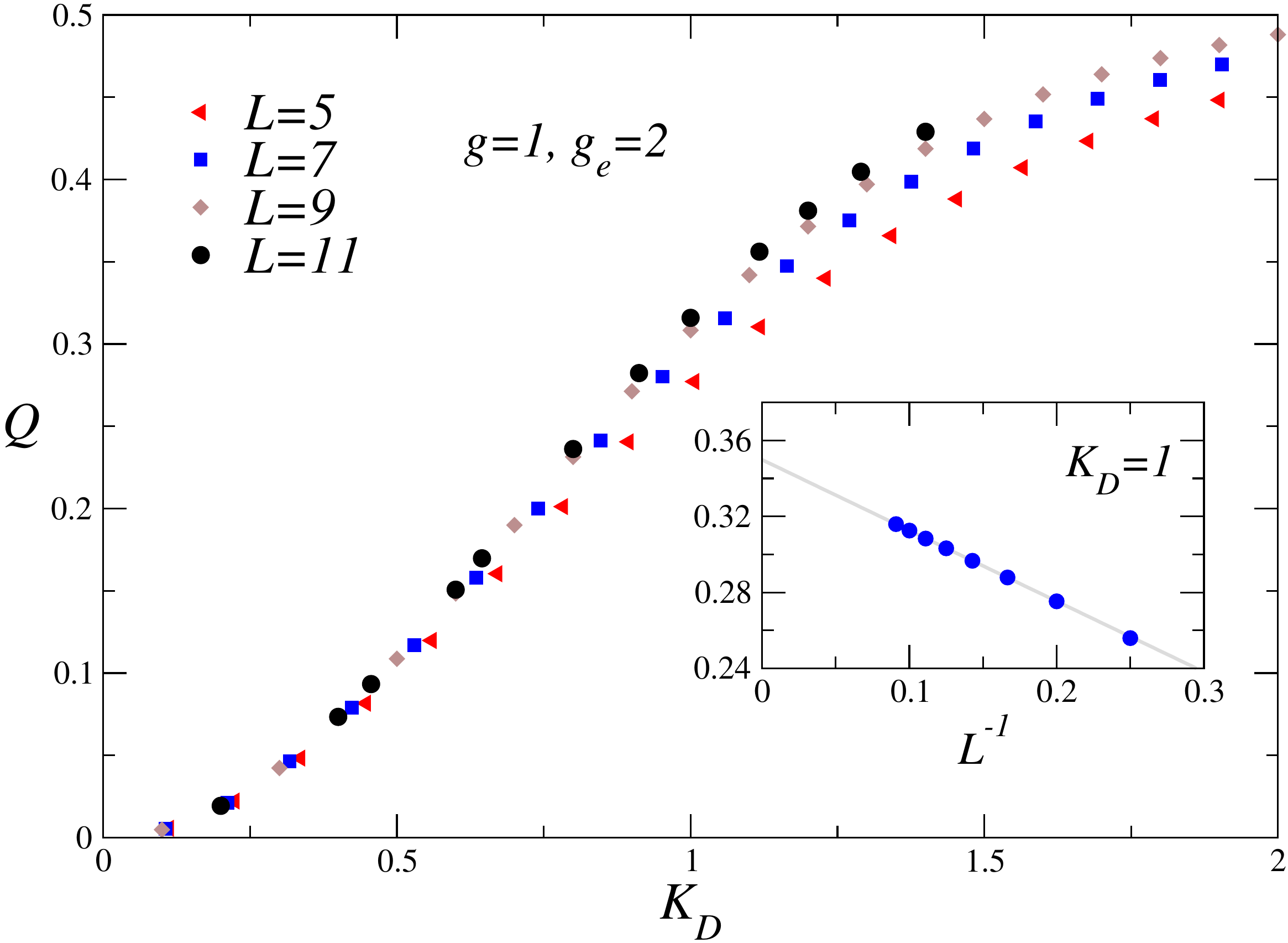}
    \caption{Scaling behavior of the decoherence factor $Q$ for the
      critical Ising chain ($g=1$), coupled with a disordered
      environment ($g_e=2$), versus $K_D=\kappa L^{1/2}$. PBC are
      used.  The results are consistent with the scaling
      relation~(\ref{tildewdef}).  The inset shows data at fixed
      $K_D=1$: size corrections decay as $L^{-1}$ (the straight line
      is only meant to guide the eye), consistently with the RG
      arguments reported in the text.}
    \label{figgcdisorderedQ}
\end{figure}

The RG irrelevance of the coupling between the two subsystems does not
imply that this coupling is negligible. First, the coupling gives rise
to a shift of the critical point $g_c(\kappa)$, see the next
subsection.  Moreover, it implies that, along the line $r=0$ and for
small $\kappa$, $R$ scales as (we assume $\kappa\ge 0$)
\begin{equation}
R \approx F_c(K_D)\,,\qquad K_D \equiv \kappa L^{y_r/2} \,.
  \label{tildewdef}
\end{equation}
The scaling behavior of the decoherence susceptibility
defined in Eq.~(\ref{chiqdef}) can be derived by
differentiating the scaling equation  $Q
\approx {\cal Q}(W_s)$ with respect to $\kappa$. We obtain
\begin{eqnarray}
  \chi_Q(g,g_e,L) \approx L^{y_r} {\cal C}(r L^{y_r})\,.
  \label{chiqscadis}
\end{eqnarray}
The power-law divergence of $\chi_Q$ shows that the coherence
properties of ${\cal S}$ are strongly affected by the coupling with
the environment.

The previous scaling relations hold modulo scaling corrections that
vanish as $L\to \infty$. They are expected to be analogous to those
arising at the critical point of isolated Ising chains. They depend on
the observable and on a variety of sources, such as irrelevant
operators, analytic backgrounds, analytic expansions of the scaling
fields, the presence of boundaries, etc.; see, for example,
Ref.~\cite{CPV-14} for a thorough discussion of this point.  In
particular, in the presence of boundaries (for instance, when OBC are
used), one expects boundary-related corrections decaying as $L^{-1}$.
These corrections are absent when PBC are used~\cite{CPV-14}.
However, we note that, at criticality ($r=0$), the corrections due to
the terms of order $\kappa^4$ in the expansion (\ref{ug}) of the
scaling field $u_r$ may become the most relevant ones. If $u_r \approx
b\,\kappa^2 + b_2\,\kappa^4$ for $r=0$, we have $W_s \approx b\, K_D^2
+ b_2 K_D^4 L^{-y_r}$, which shows that these terms contribute
corrections of order $L^{-y_r}$, at fixed $K_D=\kappa L^{y_r/2}$.  For
one-dimensional Ising systems with PBC, these corrections, of order
$1/L$, are the leading ones for the decoherence factor $Q$.  On the
other hand, the scaling corrections for $R_\xi$ are still dominated by
the analytic background~\cite{CPV-14}: they decay as $L^{-3/4}$.

\begin{figure}
 \centering
    \includegraphics[width=0.95\columnwidth]{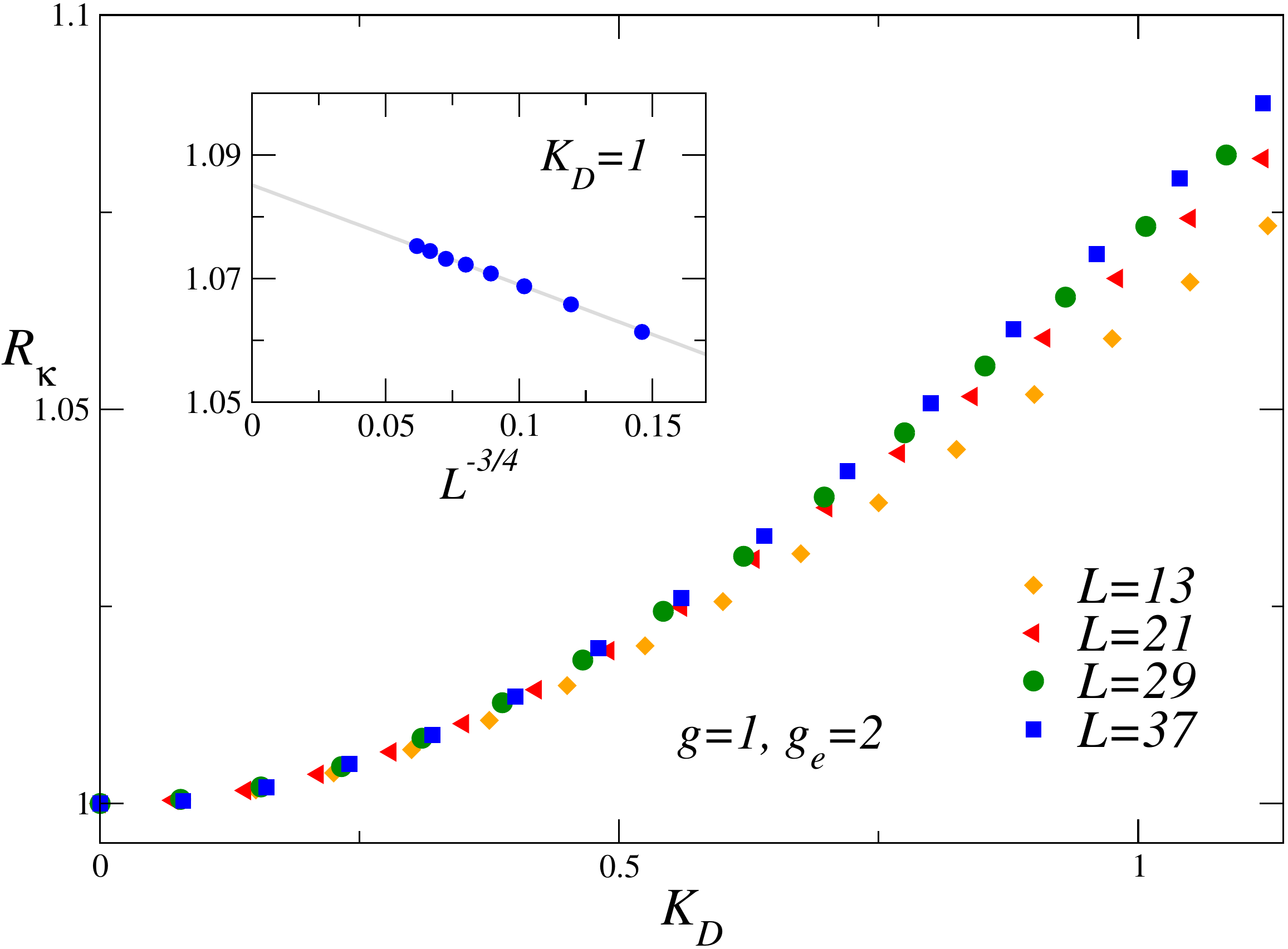}
    \caption{Plot of the ratio $R_\kappa$, defined in
      Eq.~(\ref{xirdef}), versus $K_D=\kappa L^{1/2}$, for $g=1$ and
      $g_e=2$.  Data appear to converge with increasing $L$ onto a
      single curve, supporting the scaling behavior,
      Eq.~(\ref{tildewdef}).  In the inset we show data for $K_D=1$:
      scaling corrections apparently behave as $L^{-3/4}$ (the
      straight line is only meant to guide the eye), as predicted by
      RG arguments.}
    \label{figgcdisorderedxi}
\end{figure}

To verify the previous scaling predictions, we present numerical
results for one-dimensional stacked Ising chains. We fix $g = 1$ and a
value $g_e>1$, so that the system $\cal S$ is critical and the
environment $\cal E$ is disordered for $\kappa = 0$.  In
Fig.~\ref{figgcdisorderedQ} we show data for the decoherence factor
$Q$ for systems with PBC up to $L=11$.  The results are consistent
with the RG prediction, Eq.~(\ref{tildewdef}), which, in turn, implies
Eq.~(\ref{chiqscadis}), i.e., the divergence of the decoherence
susceptibility as $L^{y_r}$.  Also the behavior of the scaling
corrections---they decay as $1/L$---is consistent with the general
theory (see the inset).  We also computed the ratio $R_\xi\equiv
\xi/L$ for systems with OBC up to $L\approx 40$.  In
Fig.~\ref{figgcdisorderedxi} we plot the ratio
\begin{equation}
  R_\kappa(g,g_e,\kappa,L) \equiv {R_\xi(g,g_e,\kappa,L)\over
    R_\xi(g=1,g_e,\kappa=0,L)}\,,
  \label{xirdef}
\end{equation}
which should scale as reported in Eq.~(\ref{Rscal}) or, for
$g=1$, as in Eq.~(\ref{tildewdef}).  The ratio $R_\kappa$ is
particularly convenient because its scaling corrections turn out to be
significantly smaller than those affecting $R_\xi$.  Again results
nicely support the RG predictions.  Also the scaling corrections, see
the inset, are consistent with the RG theory.

\subsubsection{Ising-like transition lines for small $\kappa$}
\label{disordtranline}

As already anticipated, we may also predict the behavior of the Ising
transition line $g_c(\kappa)$, starting  at the critical point 
$g_c = g_{\cal I}$ for $\kappa=0$. Transitions occur on the line
$u_r=0$. Eq.~(\ref{ug}) implies 
\begin{equation}
  g_c(g_e,\kappa) - g_c(g_e,\kappa=0) = g_c(g_e,\kappa) - g_{\cal I}
  \approx b(g_e)\,\kappa^2\,,
  \label{gcbeh}
\end{equation}
where $b$ depends only on $g_e$.  The behavior (\ref{gcbeh}) holds for
sufficiently small values of $\kappa$, when higher-order $O(\kappa^4)$
terms in the expansion (\ref{ug}) can be neglected.

In App.~\ref{AppB} we determine $b(g_e)$ for large values of $g_e$,
obtaining 
\begin{equation}
  b(g_e) \approx {3\over 2 g_e^2} \qquad {\rm for} \; g_e\to \infty\,.
  \label{blgbe}
\end{equation}
The vanishing of $b(g_e)$ far large $g_e$ follows from the more 
general result $g_c(g_e,\kappa) = g_{\cal I}$ for any $\kappa$ in
the limit $g_e\to \infty$. Indeed, in this limit the ground state of the global
system  is an eigenvector of $\tau_{\bm x}^{(3)}$ with eigenvalue 1
for all lattice points $\bm x$.  The global ground state is therefore 
factorized, i.e., 
$|\Psi_0\rangle = | \phi_0\rangle_{\cal E} \otimes |
\psi_0\rangle_{\cal S}$, where 
$| \phi_0\rangle_{\cal E} = \Pi_{\bm x}|+\rangle_{\bm x}$, 
$|+\rangle_{\bm x}$ is the $+1$ eigenvector of $\tau_{\bm x}^{(3)}$, 
and $|\psi_0\rangle_{\cal S}$ is defined on 
$\cal S$ only. Since the matrix element of $H_{\cal SE}$ on this 
factorized state vanishes, $|\psi_0\rangle_{\cal S}$ is the 
ground state of an isolated single Ising chain, that has critical point 
for $g=g_{\cal I}$, independently of $\kappa$.

We now consider the opposite limit, $r_e\equiv g_e-g_{\cal I}\to
0$. In this case the coefficient $b(g_e)$ is expected to diverge as 
$b(g_e) \sim r_e^{-\zeta}$. with $\zeta>0$. The exponent
$\zeta$ will be determined in Sec.~\ref{traline}, by matching
Eq.~(\ref{gcbeh}) with the asymptotic multicritical behavior arising
when also the environment is critical. This allows us to obtain the
exponent $\zeta$ in terms of the Ising critical exponents. We find  $\zeta
= 2 (2-\eta)\nu-1>0$ and, in particular, $\zeta = 5/2$ for one-dimensional
stacked Ising chains. The divergence of $b(g_e)$ for $r_e\to 0$,
indicates that a different regime emerges when the environment
${\cal E}$ is critical, as it will be discussed in Sec.~\ref{critenv}.

\begin{figure}
    \centering
    \includegraphics[width=0.95\columnwidth]{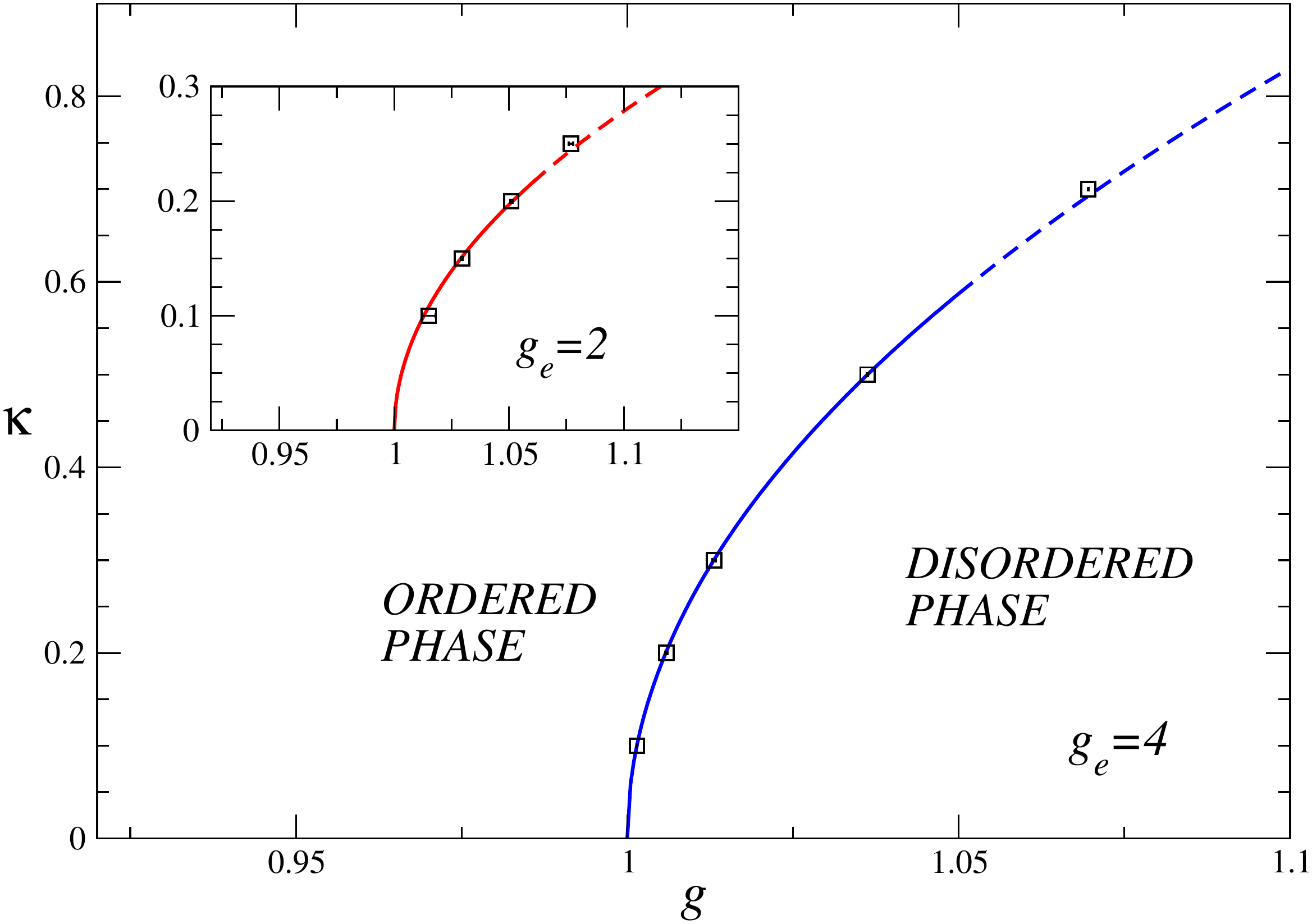}
    \caption{The $g$-$\kappa$ phase diagram when the environment is
      disordered in the absence of coupling, i.e., for $g_e>g_{\cal
        I}=1$. We report the estimates of the critical points for
      $g_e=4$ and some values of $\kappa$. In the inset we report
      analogous data for $g_e=2$.  The uncertainty on the estimates is
      smaller than, or at the most of the order of, the size of the
      symbols.  The critical lines are obtained by fitting the data
      for the smallest values of $\kappa$ (those along the full lines)
      to $g_c(\kappa)=1 + b\,\kappa^2$, obtaining $b \approx 0.14$ for
      $g_e=4$, and $b \approx 1.29$ gor $g_e=2$.  These results fully
      support the RG prediction, Eq.~(\ref{gcbeh}). 
      The coefficient $b$ rapidly increases when $g_e$
      approaches the critical value $g_{\cal I}=1$, consistently with 
      the asymptotic behavior $b(g_e)\sim (g_e-1)^{-5/2}$,
      see Eq.~(\ref{bgeasy}).}
    \label{phasediagramDIS}
\end{figure}

The prediction~(\ref{gcbeh}) is nicely confirmed by the numerical
results. In Fig.~\ref{phasediagramDIS} we report the critical points
$g_c(\kappa)$ for $g_e=2$ and $g_e=4$ and a few values of
$\kappa$. The critical points were identified by looking at the
crossing points of the ratio $R_\kappa$ defined in Eq.~(\ref{xirdef}),
for systems with OBC and lattice sizes up to $L\approx 20$.

\begin{figure}
    \centering
    \includegraphics[width=0.95\columnwidth]{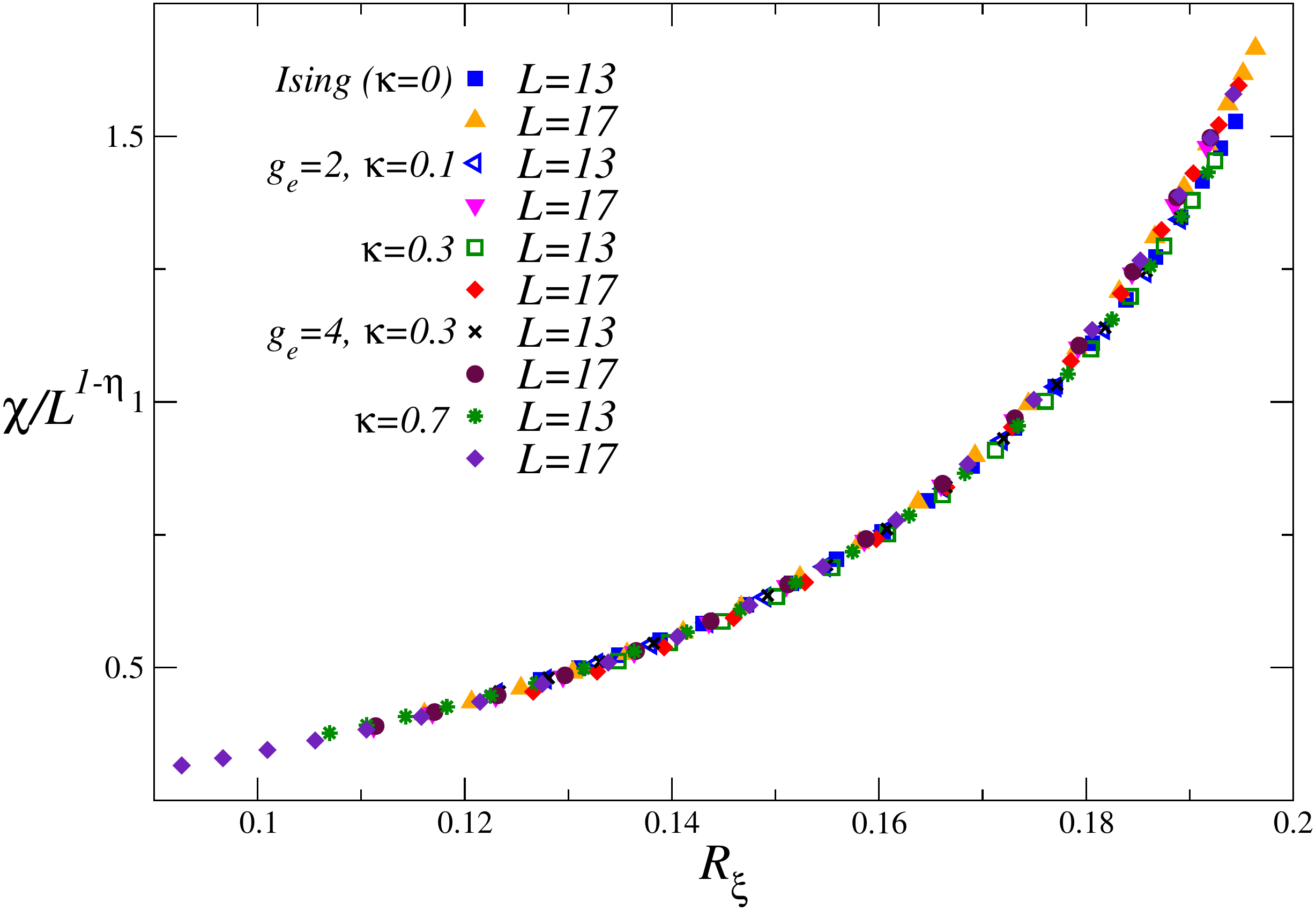}
    \caption{Plot of $\chi/L^{1-\eta}$ versus $R_\xi$.  For each $L$,
      $\kappa$, and $g_e$, data are obtained by varying $g$ around the
      critical point $g_c(\kappa)$. The Ising exponent $\eta=1/4$ is
      used.}
    \label{chiscadis}
\end{figure}

Finally, we consider the susceptibility $\chi$ as a function of
$R_\xi=\xi/L$.  In Fig.~\ref{chiscadis} we report $\chi/L^{1-\eta}$
computed varying $g$ around the critical point $g_c(\kappa)$ for
several different values of $L$ and $\kappa$, and for two values of
$g_e$.  According to the RG theory, data should scale according to
Eq.~(\ref{chiscaisi}).  The results reported in Fig.~\ref{chiscadis}
show an excellent scaling, provided we use the Ising exponent $\eta =
1/4$. They confirm that all transitions for $\kappa>0$ belong to the
2D Ising universality class. Note, moreover, that data corresponding
to different values of $g_e$ and $\kappa$ appear to approximately
collapse onto the same curve. This implies that the nonuniversal
constant $a_\chi$ is Eq.~(\ref{chiscaisi}) depends very weakly on the
system parameters.

To conclude, let us finally note that the coupling $\kappa$ also
significantly affects the properties of the environment $\cal E$: For
$\kappa \not=0$ the environment becomes critical when $\cal S$ is
driven to criticality.  Thus, only for $g > g_c(\kappa)$ is the
environment still disordered.  For $g = g_c(\kappa)$, $\cal E$ is
critical, while, for $g < g_c(\kappa)$ the environment is fully
ordered.  These features will be discussed and explained on general
grounds in Sec.~\ref{phdia}.

\subsection{Critical  environment}
\label{critenv}

We now discuss how the coupling with a {\em critical} environment
${\cal E}$ affects the critical behavior of the subsystem ${\cal S}$.
Therefore, we assume that ${\cal E}$ is close to criticality for
$\kappa = 0$, i.e., $g_e\approx g_{\cal I}$. As we shall see, the
effect of a weak coupling between ${\cal S}$ and ${\cal E}$ is
substantially different from that occurring when ${\cal S}$ is coupled
with a disordered environment.

\subsubsection{FSS at the multicritical point for $\kappa=0$}
\label{mcpointsca}

As we shall show below, if the environment ${\cal E}$ is critical, the
interaction $H_{\cal SE}$ gives rise to a relevant perturbation of the
critical behavior of the subsystem $\cal S$. In this case there are
three relevant perturbations at the uncoupled critical point
$(g=g_{\cal I},g_e = g_{\cal I},\kappa = 0)$. Two of them are those
that drive criticality in the two isolated subsystems. The coupling
$\kappa$ gives rise to an additional independent relevant RG
perturbation.

The uncoupled critical point is multicritical.  To describe the
multicritical behavior close to
it~\cite{FN-74,NKF-74,PV-02,CPV-03,BPV-22}, we introduce three
independent scaling fields,
\begin{equation}
u_r\sim r \equiv g-g_{\cal I}, \quad
u_{er}\sim r_e \equiv g_e-g_{\cal I}, \quad 
u_\kappa\sim \kappa. 
\end{equation}
Because of the equivalence of the environment and the system, $u_r$
and $u_{er}$ have the same RG dimension $y_r$. The scaling field
$u_\kappa$ has RG dimension $y_\kappa$. All scaling fields are
relevant: $y_r$ and $y_\kappa$ are both positive, see below.  Then, in
the zero-temperature and FSS limit, the singular part of the
free-energy density is expected to scale
as~\cite{Wegner-76,PV-02,CPV-03,CPV-14,RV-21}
\begin{eqnarray}
&& F_{\rm sing}(g,g_e,\kappa,L) \approx L^{-(d+z)}{\cal
    F}(W_s,W_{es},K_s) \,,
  \label{freeenL}\\
&& 
   W_s = u_r L^{y_r}\,, \quad 
   W_{es} = u_{er} L^{y_r}\,, \quad 
   K_s = u_\kappa L^{y_\kappa}\,.\qquad
\label{scalfieldcrit}
\end{eqnarray}
The scaling fields $u_r$, $u_{er}$, and $u_\kappa$ are analytic
functions of the Hamiltonian parameters $g$, $g_{e}$, and $\kappa$.
Close to the multicritical point they can be expanded as
\begin{eqnarray}
  &&u_r = r + b_\kappa \kappa^2 + b_r r^2 + b_{er} r r_e \ldots \,,
  \label{urscafi}\\
  &&u_{er} = r_e + c_\kappa \kappa^2 + c_r r_e^2 + c_{er} r r_e \ldots \,,
  \label{urescafi}\\
  &&u_\kappa = \kappa  + d_r r \kappa + d_{er} r_e \kappa \ldots \,.
  \label{ukscafi}
\end{eqnarray}
The scaling fields $u_r$ and $u_{er}$ are expected to be even under
the symmetry $\kappa\to -\kappa$ (therefore they can only depend on
$\kappa^2$), while $u_\kappa$ should be odd.  Since we are interested
in the asymptotic FSS, we may equivalently consider the simpler linear
scaling variables
\begin{equation}
  W = r L^{y_r}\,,\qquad 
  W_e = r_e L^{y_r}\,,\qquad 
  K = \kappa L^{y_\kappa}\,.
  \label{linearscalfield}
\end{equation}
This substitution is equivalent to neglecting some (typically
next-to-leading~\cite{CPV-14}, see also below) scaling corrections.
  
Close to the multicritical point, the effects of a small coupling
$\kappa$ are controlled by its RG dimension $y_\kappa$.  To compute
it, we note that the interaction term can be rewritten in the
field-theory framework as \cite{RLXP-20,LLM-98}
\begin{eqnarray}
  \int d^D x  \;\kappa \;\phi_s({\bm x}) \phi_e({\bm x})\,,\qquad D=d+z\,,
  \label{ftpert}
\end{eqnarray}
where $\phi_s$ and $\phi_e$ are the order-parameter fields for the two
critical systems ${\cal S}$ and ${\cal E}$.  Then, we
straightforwardly obtain
\begin{equation}
y_\kappa = d+z - 2 y_\phi = 2 - \eta \,,
\label{yqcomp}
\end{equation}
where $y_\phi$ is the RG dimension of the order parameter at the
$(d+1)$-dimensional Ising fixed point.  These relations give
$y_\kappa=7/4$ for $d=1$, $y_\kappa=1.963702(2)$ for $d=2$, and
$y_\kappa=2$ for $d=3$, confirming that the coupling $\kappa$
gives rise to a relevant perturbation, as anticipated above.

On the basis of the above RG analysis, we expect any RG
invariant quantity $R$ defined on the subsystem ${\cal S}$, such as
$R_\xi=\xi/L$ or the decoherence factor $Q$, to behave as
\begin{equation}
  R(g,g_e,\kappa,L) \approx {\cal R}(W,W_e,K)\,,
      \label{critsca}
\end{equation}
where ${\cal R}$ is a universal scaling function of its arguments.
In particular, if we set $r=r_e=0$, therefore moving along the line
$g=g_e = g_{\cal I}$, Eq.~(\ref{critsca}) predicts 
\begin{equation}
  R(g=g_{\cal I},g_e=g_{\cal I},\kappa,L) \approx {\cal R}_0(K)\,.
    \label{critsca2}
\end{equation} 

Differentiating twice Eq.~(\ref{critsca}) with respect to $\kappa$,
and then setting $\kappa=0$, we obtain the leading FSS behavior of the
decoherence susceptibility $\chi_Q$ defined in Eq.~(\ref{chiqdef}):
\begin{eqnarray}
  \chi_Q(g,g_e,L) \approx L^{2y_\kappa} {\cal C}(W,W_e)\, .
  \label{chiqscacrit}
\end{eqnarray}
Again this demonstrates that the coherence properties of the quantum
critical behavior of ${\cal S}$ are very sensitive to the coupling
with the environment ${\cal E}$.  We may compare the power-law
behavior $\chi_Q(g,g_e,L) \approx L^{2y_\kappa}$ with that obtained
when ${\cal E}$ is disordered for $\kappa=0$, i.e., $\chi_Q \sim
L^{y_r}$, cf.~Eq.~(\ref{chiqscadis}).  The exponent $2 y_\kappa$ is
significantly larger than $y_r$ in any dimension (indeed $y_r=1$ and
$2 y_\kappa=7/2$ in $d=1$; $y_r\approx 1.587$ and $2y_\kappa\approx
3.928$ in $d=2$; $y_r=2$ and $y_\kappa=4$ in $d=3$) and thus, not
surprisingly, the decoherence rate for a critical environment is much
larger than that for a disordered environment.

\begin{figure}
    \centering
    \includegraphics[width=0.95\columnwidth]{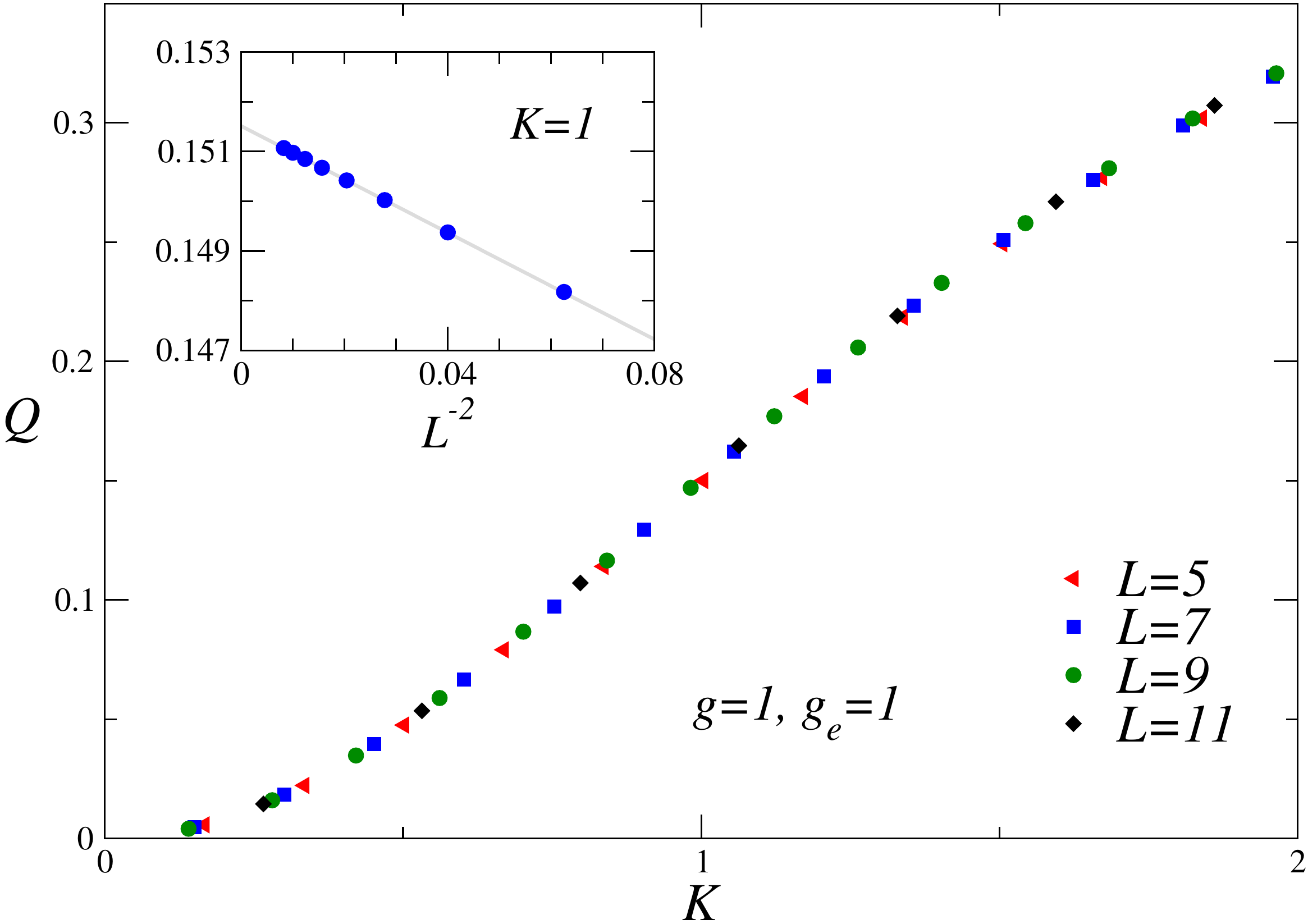}
    \caption{The decoherence function $Q$ for $g=g_e=1$
      (correspondingly, $r=r_e=0$).  The data show an excellent FSS
      when plotted versus $K=\kappa L^{7/4}$, confirming the RG
      analysis.  The inset shows the data at fixed $K=1$: size
      corrections apparently decay as $L^{-2}$, as predicted by the RG
      arguments (the line is only drawn to guide the eye).}
    \label{figdecoherencecritical}
\end{figure}

\begin{figure}
    \centering
    \includegraphics[width=0.95\columnwidth]{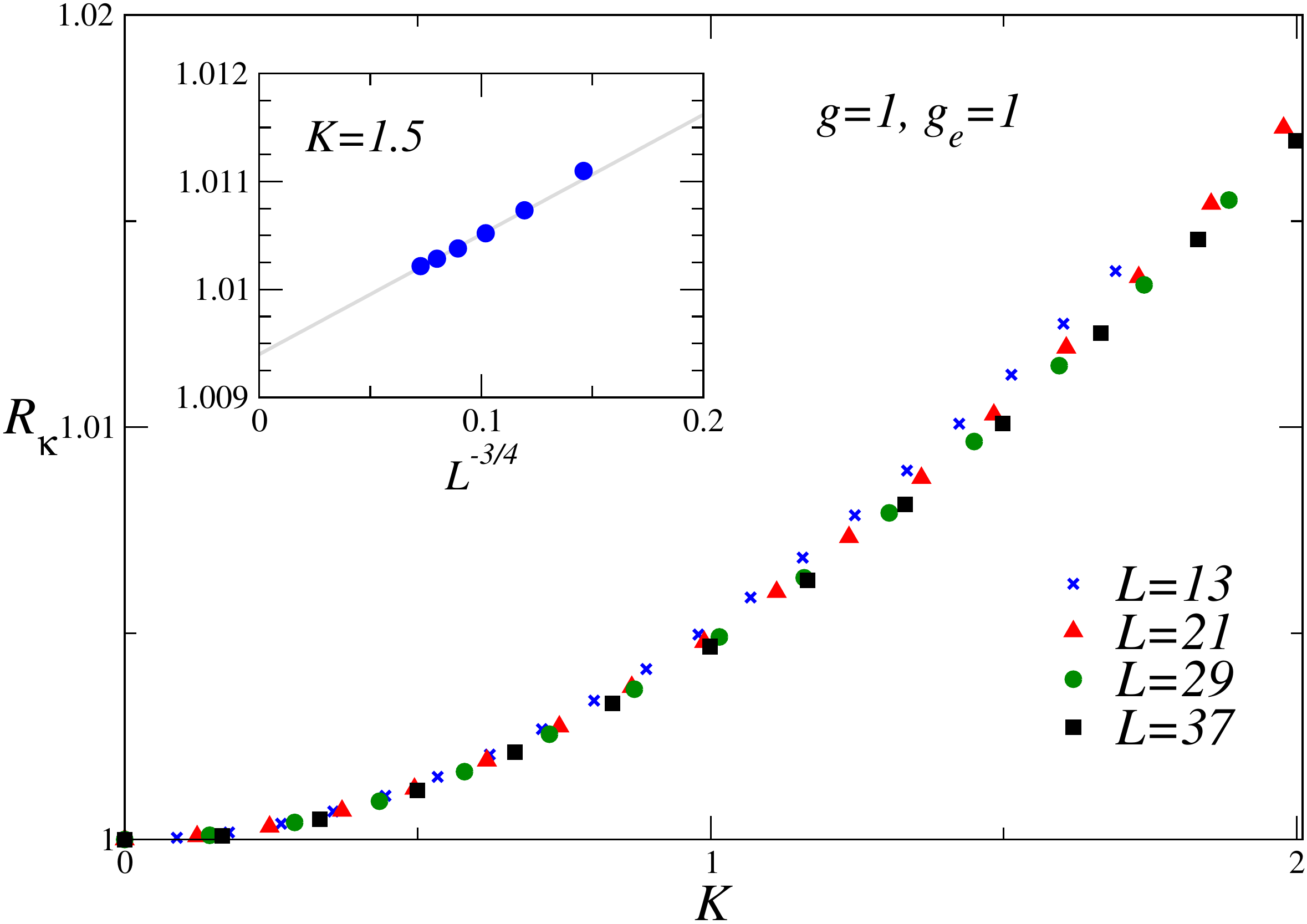}
    \caption{Scaling behavior of the ratio $R_\kappa$, defined in
      Eq.~(\ref{xirdef}), for $g=g_e=1$ (corresponding to
      $r=r_e=0$). The data show an asymptotic FSS when plotted versus
      $K=\kappa L^{7/4}$, confirming the RG analysis.  The inset shows
      the data at fixed $K=1.5$: size corrections decay as $L^{-3/4}$,
      as expected from the RG arguments (the line is only drawn to
      guide the eye).}
    \label{figxixi0criticalr0}
\end{figure}

To verify the previous scaling predictions, we have considered two
coupled chains at criticality, i.e., for $g=g_e=g_{\cal I}=1$.  We
recall that $y_r=1$ and $y_\kappa=7/4$ for one-dimensional systems.
Figs.~\ref{figdecoherencecritical} and \ref{figxixi0criticalr0} show
the behavior of the decoherence factor $Q$ (we use PBC) and of the
ratio $R_\kappa$ defined in Eq.~(\ref{xirdef}) (we use OBC),
respectively.  The results for both $Q$ and $R_\kappa$ nicely support
the scaling behavior, Eq.~(\ref{critsca2}).  In particular, the
observed scaling behavior of $Q$ implies the divergence of the
decoherence susceptibility, $\chi_Q\sim L^{2y_\kappa}$.

Scaling corrections are expected to be similar to those arising in the
case of an isolated critical Ising chain, see also Sec.~\ref{disenv}.
For one-dimensional quantum Ising systems, the leading scaling
correction for the ratio $R_\xi=\xi/L$ is expected to be due to the
analytic background~\cite{CPV-14}: it should decay as $L^{-3/4}$ for
both PBC and OBC. The scaling corrections associated with $Q$ are
expected to decay faster, as $L^{-2}$ in the absence of boundaries
(for instance, for PBC), and as $L^{-1}$ for systems with boundaries
(in the OBC case).  The insets of Figs.~\ref{figdecoherencecritical}
and \ref{figxixi0criticalr0} show that scaling corrections behave as
predicted by the RG arguments.

We remark that similar multicritical behaviors should also emerge when
the two subsystems ${\cal S}$ and ${\cal E}$ are different.  The
multicritical fixed point is given by the two decoupled fixed points
associated with the critical behaviors of ${\cal S}$ and ${\cal E}$ in
the absence of any coupling.  The RG dimension of the parameter
$\kappa$ that parametrizes the coupling of the two subsystems can be
computed again using Eq.~(\ref{ftpert}), where $\phi_{s}$ and $\phi_e$
represent the operators defined in ${\cal S}$ and ${\cal E}$ entering
the interaction Hamiltonian $H_{\cal SE}$.

\subsubsection{Ising transition lines for small $\kappa$}
\label{traline}

So far we have considered the behavior around the multicritical
point. In the parameter space $(g,g_e,\kappa)$, the multicritical
point $(g=g_{\cal I},g_e=g_{\cal I},\kappa=0)$ belongs to a
two-dimensional surface of critical transitions that lies, for $\kappa
\not=0$, in the region $g>g_{\cal I}$, $g_e > g_{\cal I}$. Such
transitions are related to the spontaneous breaking of the residual
${\mathbb Z}_2$ symmetry of the global system when $\kappa\neq 0$.
Therefore, they are expected to belong to the $(d+1)$-dimensional
Ising universality class. A more general discussion will be presented
in Sec.~\ref{phdia}. Here, we wish to discuss the shape of the
critical surface close to the multicritical point.

To make the discussion simple, let us consider a plane in the
parameter space such that the ratio $s\equiv r_e/r$ is constant, and
such that it intersects the transition surface along a line.  The
scaling behavior~(\ref{freeenL}) of the free energy also determines
the behavior of the transition line close to the multicritical point,
see, e.g., Refs.~\cite{FN-74,NKF-74,CPV-03}.  As $r_e/r=W_e/W$ is kept
constant, we can neglect the scaling field $W_e$, and we can rewrite
the singular free-energy density as
\begin{eqnarray}
  &&F \approx \xi_r^{-(d+z)}\widetilde{\cal F}(X,Y) \,,\label{freeenxy}\\
  && X =
  \xi_r/L\,,\quad \xi_r \sim r^{-1/y_r}\,, \quad Y =
  r^{-y_\kappa/y_r}\kappa\,, \qquad
\label{xysca}
\end{eqnarray}
where $\xi_r$ plays the role of a critical length scale.  In the
large-$L$ limit, i.e., for $X\to 0$, we may write
\begin{equation}
 F \approx \xi_r^{-(d+z)}{\cal F}_\infty(Y) \,.\label{freeeninf}
\end{equation}
Consistency of the phase diagram with Eq.~(\ref{freeeninf}) requires
that, for $\kappa\to 0$, the critical line $g_c(s,\kappa)$ is tangent
to the line corresponding to a fixed finite value of the scaling
variable $Y$. Therefore, for small values of $\kappa$, the above
scaling arguments predict that
\begin{equation}
  g_c(s,\kappa)- g_{\cal I} \approx w(s)
  \,\kappa^{\varepsilon}\,,\quad \varepsilon = {y_r\over y_\kappa}\le 1\,,
  \quad s={r_e\over r}\,,
  \label{critline}
\end{equation}
where the coefficient $w(s)$ depends on the ratio
$s=r_e/r$. Substituting the known values of $y_r$ and $y_\kappa$, we
find $\varepsilon = 4/7\approx 0.5714$ for $d=1$, $\varepsilon \approx
0.8084$ for $d=2$, and $\varepsilon = 1$ for $d=3$ [in $d=3$, there
  are probably additional multiplicative logarithms in
  Eq.~(\ref{critline})]. 
The small-$\kappa$ behavior
(\ref{critline}) significantly differs from that holding for a
disordered environment, see Eq.~(\ref{gcbeh}). 

As discussed in Sec.~\ref{disordtranline}, $g_c(s,\kappa) - g_{\cal
  I}=O(\kappa^2)$ for finite $r_e\equiv g_e-g_{\cal I}>0$, thus the
coefficient $w(s)$ of the power $\kappa^\varepsilon$ (where
$\varepsilon<2$) must vanish for $s\to \infty$. The consistency of the
scaling equation (\ref{critline}) with Eq.~(\ref{gcbeh}) in the limit
$g_e\to g_{\cal I}$ allows us to predict the limiting behavior of
$b(g_e)$ appearing in Eq.~(\ref{gcbeh}) for $g_e\to g_{\cal I}$ and of
$w(s)$ for $s\to \infty$.  Indeed, for small values of $\kappa$,
Eq.~(\ref{gcbeh}) can be rewritten as
\begin{equation} 
  g_c(g_e,\kappa) - g_{\cal I} = \left\{ {b(g_e) \, [g_c(g_e,\kappa)-g_{\cal
        I}]^{2/\varepsilon-1}}\right\}^{\varepsilon/2}
  \kappa^\varepsilon\,.
\label{gcbehcrit}
\end{equation}
We require this relation to be consistent with the
scaling behavior Eq.~(\ref{critline}) for $r_e\equiv
g_e-g_{\cal I}\to 0$. This implies 
\begin{equation}
  b(g_e) \sim r_e^{-\zeta}\,,
  \quad \zeta = {2\over \varepsilon} - 1 =
  2(2-\eta)\nu - 1\,.
  \label{bgeasy}
\end{equation}
Explicitly, $\zeta = 5/2, 1.47, 1$ for $d=1,2,3$, respectively.  
Eq.~(\ref{gcbehcrit}) also predicts the behavior of $w(s)$ when
$w(s)\to 0$, i.e., for large values of $s$. We find
\begin{equation}
   w(s) \sim s^{-\rho}\,,  \qquad \rho = 1 - \varepsilon/2 >0\, .
\label{ws-larges}
\end{equation}
For $d=1,2,3$ we have  $\rho=5/7$, $\rho
\approx 0.596$, $\rho=1/2$, respectively.

We would like to stress that the above RG arguments determine the
behavior of the critical lines starting from the multicritical point,
provided it exists.  However, they do not ensure the existence of such
line for any value of $s$. Indeed, the mean-field calculations
reported in App.~\ref{AppA}, suggest that the critical lines exist
only for $s>0$. Our numerical results, see below, confirm this
prediction.

As we shall discuss in Sec.~\ref{phdia}, if the $\cal S$ correlations
are critical, also correlations defined on $\cal E$ are critical.
Using the symmetry of the system under the exchange of $\cal S$ and
$\cal E$, see Sec.~\ref{models}, we can straightforwardly obtain a
relation on the location of the (common) critical points. The
coefficient $w(s)$ in Eq.~(\ref{critline}) must satisfy the relation
\begin{equation}
  w(s^{-1}) = s\,w(s)\,.
\label{crel}
\end{equation}
This relation combined with Eq.~(\ref{ws-larges}) implies that $w(s)$
diverges as $s\to 0$, as
\begin{equation}
  w(s) \sim s^{-\varepsilon/2}\quad  {\rm for}\; s\to 0\,.
  \label{wsss}
\end{equation}
The divergence of $w(s)$ for $s\to 0$ suggests that the transition
line starting from the multicritical point disappears for $s=0$, i.e.
when $g_e=g_{\cal I}$, confirming the mean-field analysis.
This is also supported by the numerical results for the stacked
Ising chains, see below.

\begin{figure}
    \centering
    \includegraphics[width=0.95\columnwidth]{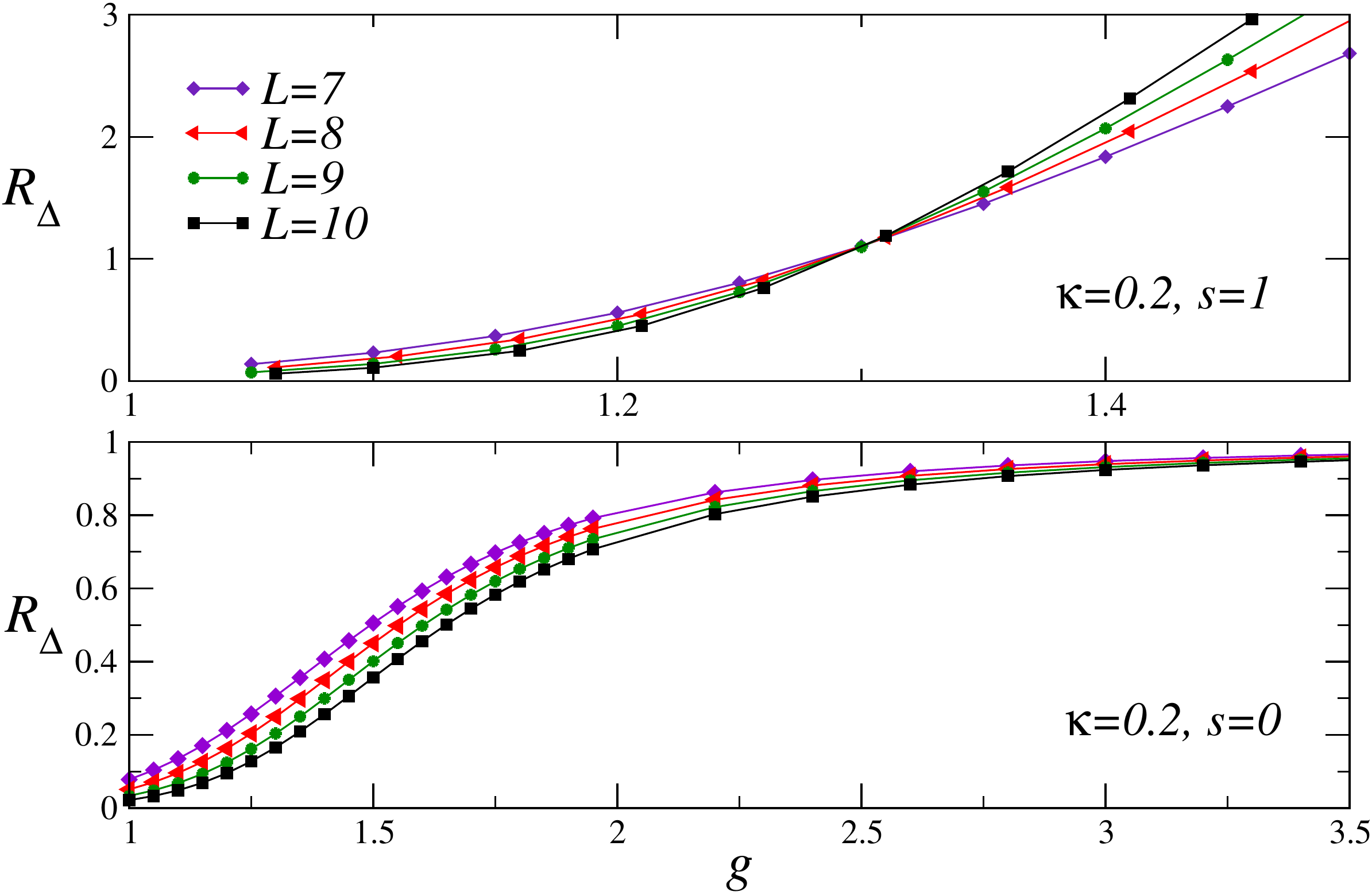}
    \caption{Plot of the ratio $R_\Delta$ defined in
      Eq.~(\ref{rdeltadef}) for $\kappa=0.2$. Top: results for $s=1$;
      bottom: results for $s=0$.  Results for $s=1$ show a clear
      crossing point for $g\approx 1.31$, indicating the presence of a
      transition. For $s=0$ the ratio $R_\Delta$ is always smaller
      than 1 and does not show any crossing, indicating that no
      transition occurs.  }
    \label{deltaratiodata}
\end{figure}

\begin{figure}
    \centering
    \includegraphics[width=0.95\columnwidth]{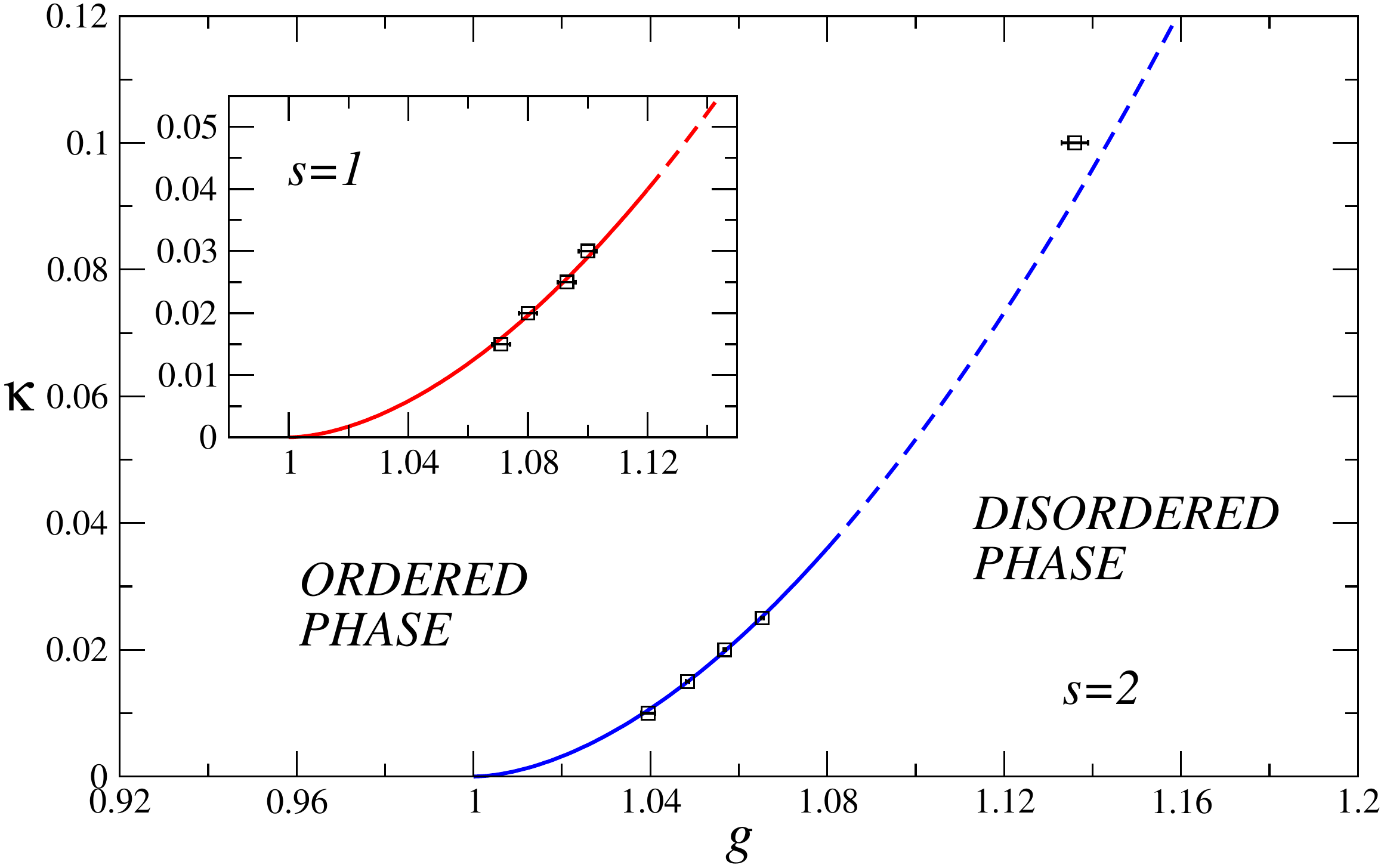}
    \caption{Estimates of the critical points $g_c(\kappa)$ for some
      values of $\kappa$ for $s\equiv r_e/r=2$ and $s=1$ (inset). The
      behavior for small $\kappa$ is in agreement with the scaling
      prediction, Eq.~(\ref{critline}).  The full lines are obtained by
      fitting the estimates of $g_c(\kappa)$ for small values of
      $\kappa$ ($\kappa\lesssim 0.03$) to $g_c(\kappa)=1 +
      w\,\kappa^\varepsilon$ with $\varepsilon=4/7$. We obtain $w
      \approx 0.53$ for $s=2$ and $w\approx 0.76$ for $s=1$.  Note
      that $w$ decreases with increasing $s$, as expected on the basis
      of the asymptotic behavior (\ref{ws-larges}).}
    \label{phasediagramcrit}
\end{figure}

To check these predictions, we again consider the stacked Ising chains
and numerically determine $g_c(s,\kappa)$ for a few small values of
$\kappa$.  For this purpose, we determine the energy gap $\Delta$,
i.e., the difference of the energies of the two lowest states of the
global system, focusing on the ratio
\begin{equation}
  R_\Delta(s,r,\kappa,L) = {\Delta(s,r,\kappa,L)\over \Delta_{\cal I}(L)}\,,
  \label{rdeltadef}
\end{equation}
where $\Delta_{\cal I}\sim L^{-1}$ is the gap of the single critical
Ising chain. We consider systems with PBC and compute the gap using
exact-diagonalization techniques. Since the transition line is
expected to belong to the Ising universality class---therefore
$z=1$---the ratio $R_\Delta$ is expected to vanish for
$g<g_c(s,\kappa)$ and to diverge for $g>g_c(s,\kappa)$. In particular,
close to the transition point, at fixed $\kappa$, it should scale as
\begin{equation}
  R_\Delta(s,g,\kappa) \approx {\cal R}_\Delta(U L^{y_r})\,,
  \qquad U = g - g_c(s,\kappa)\,,
  \label{rdeltasca}
  \end{equation}
where ${\cal R}_\Delta(x)$ is a universal scaling function, apart from
a multiplicative factor and a rescaling of its argument.  The RG
invariant ratio $R_\xi$ should scale analogously, i.e., we should have
$R_\xi(s,g,\kappa)\approx {\cal R}_\xi(U L^{y_r})$ where ${\cal
  R}_\xi$ is universal apart from a rescaling of its argument.
Therefore, the transition point $g_c(s,\kappa)$ can be determined by
looking for the crossing point of data sets for different sizes $L$.

In Fig.~\ref{deltaratiodata} we show $R_\Delta$ at fixed $\kappa=0.2$
for two values of $s$, that is, $s=1$ and $s=0$.  The data for $s=1$
show a crossing point indicating the existence of a transition point,
while those for $s=0$ are always smaller than 1 and do not show any
crossing, indicating that there is no transition for finite $\kappa$
and $s=0$ (i.e., $g_e = 1$).  Note that for $s=0$ the ratio $R_\Delta$
is expected to approach 1 for $g\to \infty$. Indeed, when
$g\to\infty$, the ground state becomes an eigenvector of all
$\sigma_{\bm x}^{(3)}$ operators. It is immediate to verify that this
implies an effective decoupling of the critical environment ${\cal E}$
(the argument is the same as the one used to discuss the limit 
$g_e\to\infty$ in Sec.~\ref{disordtranline}). 
The gap of the global system therefore converges to the gap
$\Delta_{\cal I}$ of the isolated environment.

In Fig.~\ref{phasediagramcrit} we show some results for the critical
lines starting from the multicritical point, for some values of
$s>0$. We mention that some results for identical stacked critical
Ising chains, i.e. for $g=g_e$, thus corresponding to $s=1$, were
already reported in Ref.~\cite{RLXP-20}. Our results and those
reported in Ref.~\cite{RLXP-20} nicely confirm the scaling prediction
(\ref{critline}). Moreover, our numerical results do not provide
evidence of transitions for $s\le 0$ (including also the
marginal case $s=0$, as shown above), supporting the absence of
transition lines when one of the coupling $g$ or $g_e$ is below the
Ising-chain transition point $g_{\cal I}$, as predicted by the
mean-field analysis reported in App.~\ref{AppA}.

We also mention that consistent results (not shown) have also been
obtained by analyzing the FSS behavior of $R_\xi$ or of the ratio
$R_\kappa$ defined in Eq.~(\ref{xirdef}). The DMRG results for systems
with OBC show a slower convergence to the asymptotic large-$L$
behavior, therefore leading to less precise estimates.  This can be
explained by the different behavior of the 
 size corrections for the two observables. If PBC are used, $R_\Delta$
approaches the asymptotic value with corrections that decay as 
$L^{-2}$~\cite{CPV-14}. On the other hand, for $R_\xi$ (for both PBC and OBC),
size corrections decay slower, as $L^{-3/4}$.

\subsection{Ordered environment}
\label{ordenv}

Finally, we discuss the behavior of ${\cal S}$ when the environment
${\cal E}$ is ordered and characterized by a nonzero magnetization in
the {\em thermodynamic} limit. In Ising systems, this corresponds to
choosing $g_e < g_{\cal I}$ and appropriate boundary conditions (see
below). We argue that, as in the case of a critical environment, the
interaction term $H_{\cal SE}$ is a relevant perturbation. However,
its RG dimension $\bar{y}_\kappa$ differs from that controlling the
behavior in a critical environment, since $\bar{y}_\kappa>y_\kappa$.
Moreover, no transition lines appear at finite $\kappa$; therefore for
$g_e<g_{\cal I}$, the critical point at $g=g_{\cal I}$ and $\kappa=0$
is isolated.

For a magnetized environment, the order parameter of the subsystem
${\cal E}$ is not critical, as the ground state of ${\cal E}$ is the
superposition of magnetized states with magnetization $\pm m_0$ (the
superposition must be such that the global magnetization vanishes
because of the global ${\mathbb Z}_2$ symmetry of the global
Hamiltonian).  Therefore, in the perturbation (\ref{ftpert}) we can
replace $\phi_e({\bm x})$ with an average magnetization, which is a
constant under RG transformations.  The RG perturbation (\ref{ftpert})
reduces to
\begin{equation}
  \int d^D x \;\kappa \;\phi_s({\bm x})\,,
  \label{redpert}
\end{equation}
which leads to
\begin{equation}
\bar{y}_\kappa = d+z - y_\phi = y_h\,,
\label{yqcomp2}
\end{equation}
where $y_h$ is the RG dimension of the leading symmetry-breaking
perturbation at the Ising fixed point, associated with a longitudinal
field $h$, see Sec.~\ref{models}. Therefore, $y_h = 15/8$ for $d=1$,
$y_h \approx 2.482$ for $d=2$, and $y_h = 3$ for $d=3$.  These results
imply that the singular part of the free-energy
density in the zero-temperature and FSS limit scales as
\begin{eqnarray}
&&  F_{\rm sing}(r,\kappa,L) \approx L^{-(d+z)}{\cal F}(W,K_O) \,,
  \label{freeenLord}\\
&& W \approx r L^{y_r}\,,\qquad K_O \approx \kappa
  L^{y_h}\,,
\label{scalfieldord}
\end{eqnarray}
where $r=g-g_{\cal I}$ and we used linear scaling fields.  The
specific value of the coupling $g_e<g_{\cal I}$ does not play any
role. It only changes the values of the nonuniversal constants and of
the analytic backgrounds.

The scaling behavior for any observable can be straightforwardly
obtained from those reported for the critical environment. It is
enough to replace the scaling variable $K$ with $K_O$ and $y_\kappa$
with $y_h$, in Eqs.~(\ref{critsca}), (\ref{critsca2}), and
(\ref{chiqscacrit}).  In particular, the decoherence susceptibility
behaves as
\begin{equation}
  \chi_Q\sim L^{2y_h}\,.
  \label{chiqord}
\end{equation}
Therefore, since $y_h> y_\kappa$ for any $d$ (indeed
$y_h-y_\kappa=y_\phi>0$), the sensitivity of the coherence properties
when ${\cal E}$ is in the ordered phase is even larger than the one
arising from a critical environment ${\cal E}$, where $\chi_Q\sim
L^{2y_\kappa}$.

The
leading scaling corrections for $R_\xi$ and $Q$ are analogous to those
found for a critical environment.  For example, for one-dimensional
systems with PBC, we expect $O(L^{-2})$ corrections for $Q$ and
$O(L^{-3/4})$ corrections for $R_\xi$.

\begin{figure}
    \centering
    \includegraphics[width=0.95\columnwidth]{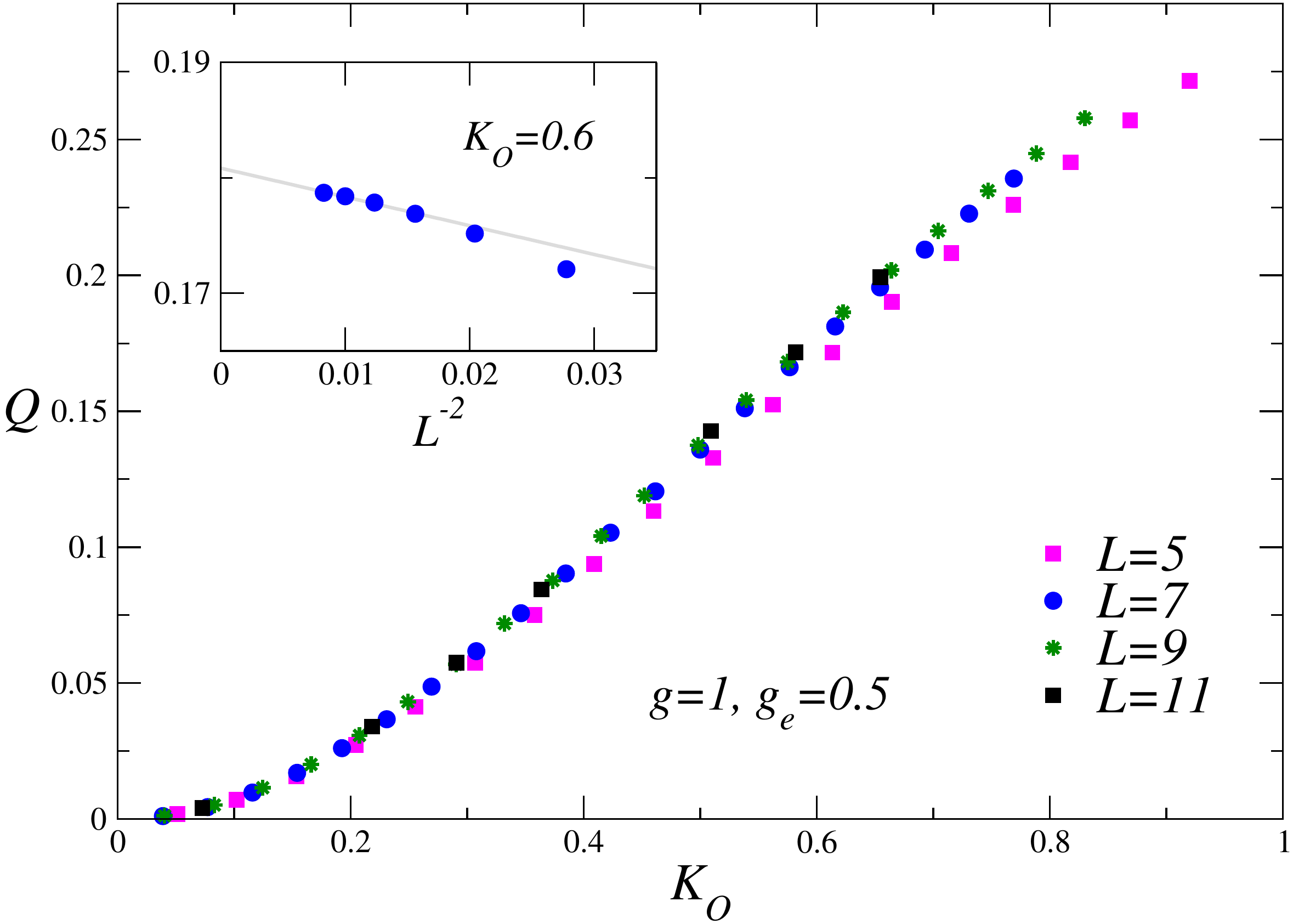}
    \caption{Decoherence function $Q$ versus $K_O=\kappa L^{y_h}$, for
      $g=g_{\cal I}=1$ and $g_e=0.5$ (ordered environment). In the
      inset, we show data for $K_O=0.6$: size corrections decay as
      $L^{-2}$ in agreement with RG arguments (the line is drawn to
      guide the eye).}
    \label{figscalingdecoherenceordered}
\end{figure}

\begin{figure}
    \centering
    \includegraphics[width=0.95\columnwidth]{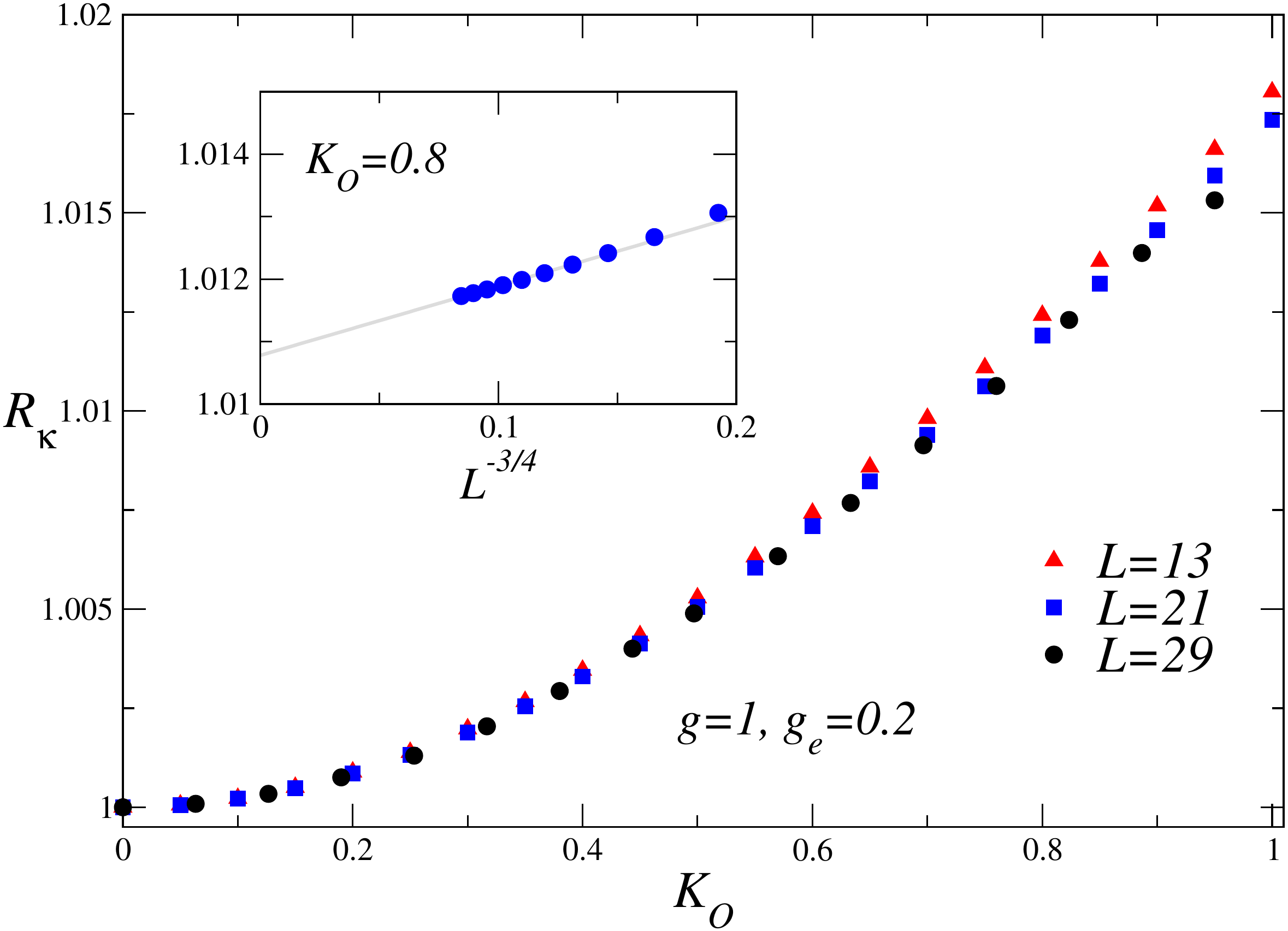}
    \caption{Ratio $R_\kappa$, defined in Eq.~(\ref{xirdef}), versus
      $K_O= \kappa L^{y_h}$, for $g=g_{\cal I}=1$ and $g_e=0.5$
      (ordered environment). In the inset, we show data at fixed $K_O
      =0.8$: size corrections decay as $L^{-3/4}$ in agreement with RG
      arguments (the line is drawn to guide the eye).}
    \label{figscalingxixi0ordered}
\end{figure}

The FSS predictions are confirmed by the results of numerical
computations for coupled quantum Ising chains.  We fixed $g=g_{\cal
  I}=1$ and considered two values of $g_e$, $g_e = 0.2$ and
$g_e=0.5$. Results for the decoherence factor $Q$ for a system with
PBC and for $R_\kappa$ defined in Eq.~(\ref{xirdef}) for a system with
OBC are shown in Figs.~\ref{figscalingdecoherenceordered} and
\ref{figscalingxixi0ordered}, respectively.  The results nicely
support the scaling behavior (\ref{freeenLord}).

Since the parameter $\kappa$ effectively behaves as an external
longitudinal field when $g_e<g_{\cal I}$, we do not expect any
transition for finite values of $\kappa$, as in the standard quantum
Ising model in the presence of an external longitudinal field
$h$. This is confirmed by numerical results.

We finally stress that the derivation of the above scaling behaviors
assumes that the environment ${\cal E}$ is subject to neutral boundary
conditions, i.e., to boundary conditions that do not favor any ordered
phase, such as PBC and OBC. Only with these boundary conditions is the
ground state a superposition of magnetized states~\cite{CNPV-14}. Some
important changes may occur when different boundary conditions are
considered, for example, fixed boundary conditions that favor one of
the broken phases, or antiperiodic boundary conditions. Indeed, the
FSS of systems at quantum first-order transitions drastically depends
on the nature of the boundary conditions, see, e.g.,
Refs.~\cite{CNPV-14,CPV-15,PRV-18-fo,RV-21}.

\section{General phase diagram} \label{phdia}

In Sec.~\ref{smallqsca}, we discussed the behavior of the subsystem
${\cal S}$ in the weak-coupling regime $\kappa \ll 1$. We wish to
discuss now the behavior for finite values of $\kappa$.  We recall
that the phase diagram of the global system must be symmetric with
respect to a change of the sign of $\kappa$ at fixed $g$ and
$g_e$. Moreover, at fixed $\kappa$, it must be symmetric with respect
to interchanging $g$ and $g_e$, due to the fact that the two
subsystems ${\cal S}$ and ${\cal E}$ are identical apart from the
transverse couplings $g$ and $g_e$.

A mean-field analysis is presented in App.~\ref{AppA}, which shows
that, also for finite $\kappa$, the environment parameter $g_e$ plays
an important role. The main features of the mean-field analysis are
substantially confirmed by the numerical results that we have obtained
for the stacked Ising chains.

If the environment is ordered in the absence of coupling, i.e., for $g_e <
g_{\cal I}$, no transition occurs for any value of $\kappa$.  The
coupling with the environment drives the system ${\cal S}$ in the
ordered phase, independently of the value of $g$.  A critical behavior is
only observed for $\kappa = 0$ and $g = g_{\cal I}$, as
discussed in Sec.~\ref{ordenv}.
A different behavior is observed for $g_e > g_{\cal I}$, i.e., when
the environment is disordered in the absence of coupling. In this
case, by tuning $g$ to a critical value $g_c(\kappa,g_e) > g_{\cal
  I}$, a critical line appears, which should be associated with the
breaking of the global ${\mathbb Z}_2$ symmetry, cf.
Eq.~(\ref{globalZ2}). This is shown in Fig.~\ref{phasediagramDIS}. 

It is interesting to observe that, at the critical point, both
correlations defined on $\cal S$ and on $\cal E$ are critical. This is
a specific feature of the interaction we consider, or, more precisely,
of the invariance properties of the interaction term $H_{\cal SE}$.
In general, at fixed $g$ and $g_e$, one may expect different types of
transitions: a transition at $\kappa = \kappa_{\cal S}(g,g_e)$, where
$\cal S$ correlations are critical, and a transition at $\kappa =
\kappa_{\cal E}(g,g_e)$, where $\cal E$ correlations are
critical. Note that we are making no assumption of the existence of
the transitions: at fixed $g$ and $g_e$, there may be no transition,
one transition, or both of them.  The symmetry of the system under the
exchange of $\cal S$ and $\cal E$ implies $\kappa_{\cal S}(g,g_e) =
\kappa_{\cal E}(g_e,g)$, but not the equality of the two
functions. The relation $\kappa_{\cal S}(g,g_e) = \kappa_{\cal
  E}(g,g_e)$ follows instead from the analysis of the symmetry
breaking pattern at the transition. Indeed, in our model the symmetry
involves transformations on both systems, see
Eq.~(\ref{globalZ2}). Thus, we expect both systems to be magnetized on
the ordered side of the transition, which, in turn, implies that they
become critical simultaneously.  This prediction has been explicitly
verified in Sec.~\ref{ordenv}. Note that the equality of the two
transition functions does not necessarily hold, if the interaction
term does not break the symmetry under independent longitudinal spin
reflections. This would be the case of subsystems coupled by using the
transverse spin operators, for instance, by the Hamiltonian interaction
(\ref{altint}). In this case, for $g\not= g_e$, one might observe two
different transitions.

Since the $\cal E$ and the $\cal S$ subsystems become simultaneously
critical, for $g < g_c(\kappa)$, both $\cal S$ and ${\cal E}$ should
be ordered, as verified in Sec.~\ref{ordenv}.  For $g > g_c(\kappa)$
instead, both subsystems should be disordered. Since the global
${\mathbb Z}_2$ symmetry is broken at $g_c(\kappa)$, the transition
should belong to the Ising universality class and $g - g_c(\kappa)$
should represent the scaling field associated with the leading even
perturbation, analogous to $r=g-g_{\cal I}$ for standard Ising
systems. This is confirmed by the scaling plots reported in
Fig.~\ref{chiscadis}.

The intermediate situation where $g_e\approx g_{\cal I}$ gives rise to
a multicritical behavior, as discussed in Sec.~\ref{critenv}. In
agreement with the mean-field analysis, the numerical results confirm
that there are no transitions at finite $\kappa$ when $g_e\le g_{\cal
  I}$.  In the opposite case $g_e > g_{\cal I}$, the transition lines
are present.  Their behavior for $\kappa\to 0$ and $g_e\to g_{\cal
  I}$, is consistent with the expected multicritical scaling, see
Fig.~\ref{phasediagramcrit}.

Finally, let us consider the behavior in the limit  $\kappa\to\infty$. 
A general discussion is reported in App.~\ref{AppB}. We show that 
the system is fully ordered for $\kappa\to\infty$ at fixed $g$ and $g_e$. 
For finite, large values of $\kappa$, a transition always occurs for large
values of $g$. More precisely, $g_c(\kappa) \approx a \kappa$ for
$\kappa\to\infty$, where $a$ is a constant dependent on $g_e$, see
App.~\ref{AppB}.

\section{Out-of-equilibrium dynamic scaling behavior}
\label{dynsca}

The recent progress achieved by quantum simulators in controlling the
dynamics of an increasing number of qubits has called for theoretical
investigations of the coherent time evolution of quantum correlations
in composite systems, of the decoherence of one subsystem due to the
interaction with the remainder, and of the energy exchanges between finite-size 
subsystems (see, e.g., Refs.~\cite{RV-21, Dziarmaga-10, PSSV-11, GAN-14,
  NC-book}).  A deeper understanding of the decoherence and
entanglement dynamics is of fundamental importance, both for
quantum-information purposes and for the improvement of energy
conversion in complex networks~\cite{NC-book, LCCLCN-13}.  Moreover,
the study of the energy storage and exchange among the different 
components  of a quantum
system is relevant for quantum-thermodynamical
purposes~\cite{GHRRS-16, VA-16}, as well as for the efficiency
optimization of recently developed quantum batteries~\cite{CPV-18}.

The study of the phase diagram and of the equilibrium scaling properties
reported in the previous sections allows one to describe the adiabatic 
slow dynamics of finite-size systems (we recall that finite-size
many-body systems are generally gapped).  In this section we extend
the analysis to out-of-equilibrium dynamic
protocols.  In particular, we wish to determine the dynamic scaling
behaviors induced by a time-dependent coupling between the subsystems
${\cal S}$ and ${\cal E}$.

We consider here the following quenching protocol. Initially, the
systems ${\cal S}$ and ${\cal E}$ are decoupled, i.e., $\kappa=0$, and
are in their ground states $|\Psi^{({\cal S})}_0\rangle$ and
$|\Psi^{({\cal E})}_0\rangle$, so that the initial many-body state
$|\Psi_0\rangle$ is given by
\begin{equation}
|\Psi_0\rangle =   |\Psi^{({\cal S})}_0\rangle
  \otimes |\Psi^{({\cal E})}_0\rangle \,.
  \label{eq_groundstate_before_quench}
\end{equation}
Then, at $t=0$, the system is suddenly driven out of equilibrium by
quenching the parameter $\kappa$ to a finite value, i.e., $\kappa$
varies instantaneously from zero to a finite value $\kappa>0$. The
initial state $|\Psi_0 \rangle$ is no longer a Hamiltonian eigenstate
and it evolves according to the Schr{\"o}dinger equation,
\begin{equation}
  |\Psi(t)\rangle = e^{-iHt}|\Psi_0\rangle\,,
  \label{eq_schroedinger}
\end{equation}
where $H$ is the total Hamiltonian (\ref{twosys}).

One can easily check that a sudden quench from $\kappa=0$ to any
$\kappa>0$ entails a vanishing average quantum work
\begin{eqnarray}
  W &=& \langle \Psi(t) |H_{\cal S}(g) + H_{\cal
    E}(g_e) + H_{\cal SE}(\kappa) | \Psi(t) \rangle  \nonumber\\
  &&-  \langle \Psi_0 |H_{\cal S}(g) + H_{\cal
    E}(g_e)| \Psi_0 \rangle\,.
\label{diffw}
  \end{eqnarray}
Indeed, since we are interested in a sudden quench at $t = 0$, and the
average energy is conserved for $t>0$, we can compute the average work
replacing $|\Psi(t)\rangle$ with $|\Psi_0 \rangle$, thus obtaining
\begin{eqnarray}
  W &=&
  \langle \Psi_0 |H_{\cal SE}(\kappa)| \Psi_0
\rangle \label{wcomp}\\
&=& -\kappa \sum_{\bm x}
\langle \Psi^{({\cal S})}_0| \sigma_{\bm x}^{(1)}|\Psi^{({\cal S})}_0\rangle
\langle \Psi^{({\cal E})}_0| \tau_{\bm x}^{(1)}|\Psi^{({\cal E})}_0\rangle
= 0\,,
\nonumber
\end{eqnarray}
where we used the fact that $\langle \Psi^{({\cal S})}_0| \sigma_{\bm
  x}^{(1)}|\Psi^{({\cal S})}_0\rangle$ and $\langle \Psi^{({\cal
    E})}_0| \tau_{\bm x}^{(1)}|\Psi^{({\cal E})}_0\rangle$ vanish due
to the ${\mathbb Z}_2$ symmetry of the Hamiltonians $H_{\cal S}$ and
$H_{\cal E}$. It is also worth mentioning that some average work $W_b$
is instead necessary to suddenly turn the coupling $\kappa$ off after
some time $t>0$, to go back to the original decoupled Hamiltonian,
because
\begin{equation}
W_b(t)= \langle \Psi(t) |H_{\cal S} + H_{\cal E}|
\Psi(t)\rangle - \langle \Psi_0 |H_{\cal S} + H_{\cal E}|
\Psi_0\rangle>0\,.
  \label{work2}
\end{equation}

After quenching at $t=0$, the energy of the global system is conserved
along the evolution for $t>0$. However, we may have some energy
exchange between the subsystems ${\cal S}$ and ${\cal E}$. This can be
quantified by the average energy exchange $E_{\rm ex}$ defined as
\begin{equation}
  E_{\rm ex}(t) = \langle \Psi(t) |H_{\cal S}| \Psi(t)\rangle
  - \langle \Psi_0
  |H_{\cal S}| \Psi_0\rangle\,.
  \label{aveeex}
\end{equation}

To monitor the coherence properties of the subsystem ${\cal S}$ along
the time evolution, one may define a time-dependent decoherence
function $Q(t)$ analogous to the equilibrium decoherence factor $Q$,
cf. Eq.~(\ref{purentdec}),
\begin{equation}
  Q(t) = 1 - {\rm Tr}\,\rho_{\cal S}(t)^2\,,\quad
  \rho_{\cal S}(t)\equiv{\rm Tr}_{\cal E}
\big[|\Psi(t)\rangle \langle \Psi(t)|\big]\,.
  \label{def_decoherence_dyn}
\end{equation}
where $\rho_{\cal S}(t)$ is the time-dependent reduced density matrix
of the system ${\cal S}$. One may also define correlations functions
at fixed time, i.e.
\begin{equation}
G(t,{\bm x},{\bm y}) \equiv {\rm Tr} \big[
  \rho_{\cal S}(t) \,\sigma_{\bm x}^{(1)} \sigma_{\bm y}^{(1)} \big]\,,
\label{gtdef}
\end{equation}
and extract a corresponding correlation length, analogously to
Eq.~(\ref{chixidef}).

One may generally distinguish between two types of sudden
quench~\cite{RV-21}: a {\em soft} quench is related to a tiny change
of the parameter $\kappa$ (decreasing with $L$), so that the system
stays close to a quantum transition and thus excites only critical
low-energy modes. In contrast, a {\em hard} quench is not limited by
the above condition and typically involves the injection into the
system of an extensive amount of energy, in such a way that also {\em
  high-energy} excitations are involved.  In the following we only
discuss soft quenches, and put forward the appropriate scaling
behaviors in a dynamic FSS framework.

We generalize the equilibrium scaling description of the weak-coupling
regime, outlined in Sec.~\ref{smallqsca}, to the out-of-equilibrium
case in which $\kappa$ varies from zero to a nonzero value, that is so
small that the evolution can be considered a soft quench.  In general,
dynamic behaviors exhibiting a nontrivial time dependence require the
introduction of another scaling variable associated with the time
variable $t$, defined as~\cite{PRV-18-qu,RV-21}
\begin{equation}
  \Theta = t \, \Delta_{\cal I}(L)\,,\qquad \Delta_{\cal I}(L) \sim L^{-z}\,, 
  \label{defTheta}
\end{equation}
where $\Delta_{\cal I}$ is the finite-size gap at criticality of the
isolated system $\cal S$, and $z$ is the dynamical exponent, which is
equal to 1 for any $d$-dimensional Ising system.  In the dynamic FSS
limit $L\to\infty$ and $g\to g_{\cal I}$, the equilibrium scaling
variables defined before (they depend on the nature of ${\cal E}$) and
the time variable $\Theta$ defined in Eq.~(\ref{defTheta}) should all
be kept constant.

For instance, if ${\cal E}$ is at criticality, the decoherence
function $Q$ obeys the dynamic FSS scaling
law~\cite{PRV-18-qu,FRV-22-2,RV-21}
\begin{equation}
  Q(t,r,r_e,\kappa,  L) \approx \mathcal{Q}(\Theta, W, W_e, K)\,,
  \label{dynFSS_decoherence}
\end{equation}
where the scaling fields $W$, $W_e$ and $K$ are defined in
Eqs.~(\ref{linearscalfield}).  This scaling ansatz generalizes
Eq.~(\ref{critsca}) to the dynamic case, by simply adding an
additional dependence on the time scaling variable $\Theta$.
Analogous relations hold for other observables, such as the ratio
$R_\xi\equiv \xi/L$ at time $t$.  Since the average energy flow
between ${\cal S}$ and ${\cal E}$, defined in Eq.~(\ref{aveeex}), is
expected to scale as the energy gap at the transition point, i.e.,
\begin{equation}
  E_{\rm ex}\sim \Delta_{\cal I}(L)\sim L^{-z}\,,
  \label{eexsca}
  \end{equation}
in the FSS limit, it should satisfy the scaling relation
\begin{equation}
  {E_{\rm ex}(t,r,r_e,\kappa, L)\over \Delta_{\cal I}(L)}
  \approx \mathcal{E}(\Theta,W, W_e, K)\,.
  \label{FSS_energy}
\end{equation}
The average work $W_b$ defined in Eq.~(\ref{work2}) should scale
analogously.

The same scaling arguments apply when ${\cal E}$ is ordered or
disordered.  It is enough to supplement the corresponding equilibrium
scaling relations with an additional dependence on the time scaling
variable $\Theta$.

\begin{figure}
    \centering
    \includegraphics[width=0.95\columnwidth]{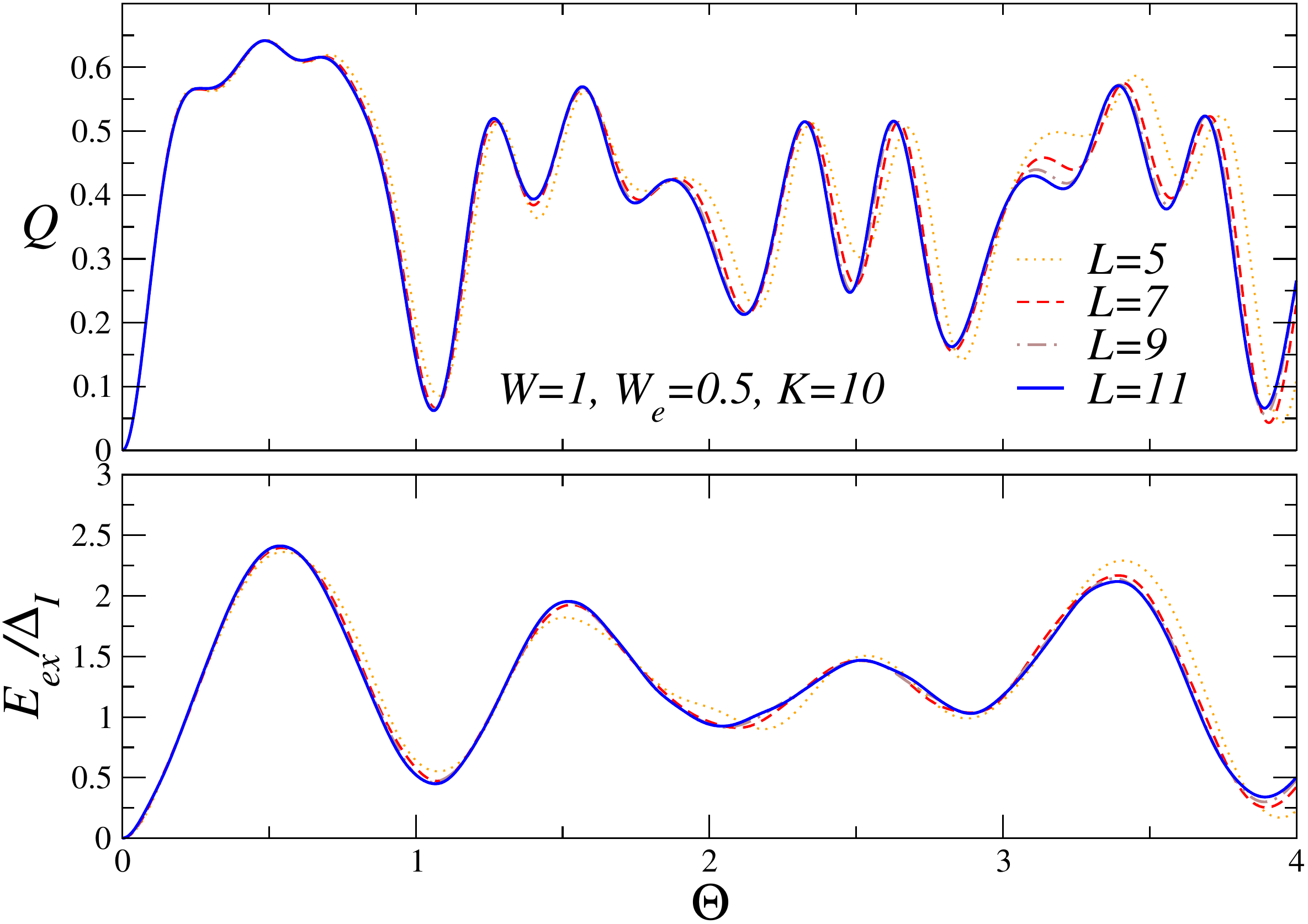}
    \caption{Plot of the ratio $E_{\rm ex}/\Delta_{\cal I}(L)$--- the
      energy exchange is defined in Eq.~(\ref{aveeex})---and of the
      decoherence factor $Q$, defined in
      Eq.~(\ref{def_decoherence_dyn}), versus $\Theta$. $\Delta_{\cal
        I}(L)\sim L^{-z}$ is the gap of the Ising chain (\ref{HSdef})
      at $g=g_{\cal I}$. For each $L$, we fix $g$, $g_e$, and
      $\kappa$, so that the equilibrium scaling variables defined in
      Eq.~(\ref{linearscalfield}) take the values $W=1$, $W_e=0.5$,
      and $K=10$ (therefore, as $L$ increases, $g,g_e\to 1$ and
      $\kappa \to 0$).  The collapse of the results for different
      sizes on a single curve supports the dynamic FSS relations
      Eqs.~(\ref{dynFSS_decoherence}) and (\ref{FSS_energy}). }
    \label{dynscafig}
\end{figure}

The above FSS predictions have been checked numerically.  We consider
stacked Ising chains with PBC and compute the ground
state~(\ref{eq_schroedinger}) by means of $4^{{\rm th}}$-order
Runge-Kutta algorithms.  The energy exchange $E_{\rm ex}$ defined in
Eq.~(\ref{aveeex}) and the decoherence factor $Q$ defined in
Eq.~(\ref{def_decoherence_dyn}) are plotted versus $\Theta$ in
Fig.~\ref{dynscafig}, close to the critical point $g=g_e=1$ (critical
environment).  The scaling behavior of the data clearly supports the
dynamic FSS predictions, Eqs.~(\ref{dynFSS_decoherence}) and
(\ref{FSS_energy}). We note that the asymptotic FSS is observed for
relatively small sizes of the system, already with 20 qubits.
Analogous results can be obtained for disordered or ordered
environments.

In conclusion, the out-of-equilibrium behavior of the subsystem ${\cal
  S}$ satisfies scaling laws that are extensions of the equilibrium
FSS relations, with the crucial addition of the scaling variable
$\Theta= t \, \Delta_{\cal I} \sim t \,L^{-z} $, where $\Delta_{\cal
  I}$ is the gap of the critical Ising system.  Similar dynamic
scaling relations hold for other slow out-of-equilibrium protocols, in
which the Hamiltonian parameters are slowly varied moving the
subsystem ${\cal S}$ across the quantum critical point point. Another
interesting case is the so-called Kibble-Zurek
dynamics~\cite{Kibble-80,Zurek-85,Zurek-96}, for which peculiar
out-of-equilibrium scaling behaviors emerge both in the
infinite-volume and in the FSS limit~\cite{CEGS-12,RV-21,TV-22}.

\section{Conclusions}
\label{conclu}

We have discussed the quantum behavior of an {\em open} many-body
system ${\cal S}$ interacting with a surrounding many-body environment
${\cal E}$, assuming that the global system is a pure state
$|\Psi_{{\cal S}\oplus {\cal E}}(t)\rangle$ that evolves unitarily.
As a paradigmatic model, we consider two coupled one-dimensional Ising
chains with Hamiltonian (\ref{twosys}), see Fig.~\ref{sketchsystem},
but the theoretical predictions apply to general $d$-dimensional Ising
systems. One of the chains plays the role of the open system ${\cal
  S}$ under observation, while the other one plays the role of
environment ${\cal E}$. The two chains interact by means of the
Hamiltonian term $H_{\cal SE}$ defined in Eq.~(\ref{HSEdef}), that
couples the longitudinal spin variables of both subsystems.  We
analyze the decoherence rate and the quantum critical behavior of the
subsystem ${\cal S}$, assuming that the global system is in its ground
state.  We also discuss the out-of-equilibrium behavior after a sudden
quench of the interaction between ${\cal S}$ and ${\cal E}$, assuming
that the global system is isolated and evolves unitarily. In
particular, we study how these equilibrium and out-of-equilibrium
behaviors depend on the quantum phases of ${\cal E}$, when ${\cal S}$
and ${\cal E}$ are weakly coupled. A detailed RG analysis shows that
three qualitatively different behaviors are observed depending on $
{\cal E}$, whether it is disordered (therefore characterized by
short-ranged correlations), critical (correlations are algebraically
decaying) or ordered (long-range correlations).  The different phases
of ${\cal E}$ give rise to different FSS behaviors with respect to the
coupling parameter $\kappa$ between ${\cal S}$ and ${\cal E}$

To quantify the effects of the interaction with the environment on the
coherence properties of ${\cal S}$, we consider the susceptibility
$\chi_Q$ of the decoherence factor $Q$, defined in
Eq.~(\ref{chiqdef}).  The susceptibility $\chi_Q$ provides a measure
of the sensitivity of the coherence properties of ${\cal S}$ to the
coupling with ${\cal E}$ for small value of $\kappa$.  It shows a
power-law divergence in the size $L$ of the system, for any state of
${\cal E}$. However, the power-law depends on the phase: $\chi_Q\sim
L^{y_r}$ when ${\cal E}$ is disordered, $\chi_Q\sim L^{2y_\kappa}$
with $y_\kappa=y_h-y_\phi$ when ${\cal E}$ is critical, $\chi_Q\sim
L^{2y_h}$ when ${\cal E}$ is ordered, where $y_r$ and $y_h$ are the RG
dimensions of the leading even and odd relevant perturbations at the
$(d+1)$-dimensional Ising fixed point, and $y_\phi$ is the RG
dimension of the order-parameter field.  Note that, since
$y_r<2y_\kappa<2y_h$ in any dimension, the decoherence rate of ${\cal
  S}$ becomes larger and larger, moving from a disordered to an
ordered environment ${\cal E}$.  For example, for one-dimensional
systems we have $y_r=1$, $2y_\kappa = 7/2$ and $2y_h=15/4$.  Numerical
results for coupled Ising chains, obtained by means of
exact-diagonalization and DMRG computations, nicely support the
scaling behaviors put forward in this paper.

We have also discussed how the equilibrium scaling behaviors can be
extended to out-of-equilibrium dynamic processes, for instance, to
protocols entailing a sudden quench of the coupling between ${\cal S}$
and ${\cal E}$, see Sec.~\ref{dynsca}.  In particular, for a {\em
  soft} quench we conjecture out-equilibrium FSS laws that extend
those valid at equilibrium. They are obtained by simply adding an
additional dependence on the time scaling variable $\Theta = t
\Delta_{\cal I}(L)$ (where $\Delta_{\cal I}\sim L^{-z}$ is the gap at
the critical point of a quantum Ising system) to the equilibrium FSS
relations.

In our analysis we focused on the behavior of ${\cal S}$ and ${\cal
  E}$, always assuming that the subsystem ${\cal S}$ under observation
is close to criticality. It would be interesting to investigate the
same issues when the subsystem ${\cal S}$ is the first-order
transition region, i.e., for $g<g_{\cal I}$. In this regime, FSS
behaviors emerge as well, although they turn out to significantly
depend on the nature of the boundary conditions, see, e.g.,
Refs.~\cite{CNPV-14,CPV-15,PRV-18-fo,PRV-18-qu,PRV-18-def,PRV-20,RV-21}.

We stress that the scenario emerging from the study of two coupled
Ising chains is expected to be quite general.  Therefore, it should be
straightforward to extend the analysis presented in this paper to
systems ${\cal S}$ and ${\cal E}$ of different nature, different
dimensionality, etc...  In particular, we expect the behavior of
${\cal S}$, including the coherence properties, phase diagram and
critical behavior, to be strongly dependent on the quantum phase of
the environment ${\cal E}$. However, we note that our general scaling
framework at the multicritical point, when both ${\cal S}$ and ${\cal
  E}$ develop critical modes (see Sec.~\ref{critenv}), was essentially
assuming a competition of critical modes characterized by equal
dynamic exponents $z$, in particular $z=1$ for both subsystem
criticalities. We believe that the competition of critical modes
associated with different dynamic exponents may lead to further
interesting features, worth being investigated.

The effects of the interactions with an environment (bath) on quantum
many-body systems have been addressed, exploiting different
approaches, in, e.g., Refs.~\cite{CL-83,LCD-87,WTS-04,SWT-04,
  WVTC-05,YMZ-14,ARBA-17,KMSFR-17,WSBR-18,NRV-19,RV-20,RV-21}.  One
approach is based on the so-called Lindblad master
equation~\cite{BP-book, RH-book}, which allows for some types of
dissipative interactions without the necessity of keeping track of the
full environment dynamics, within some approximations, see, e.g.,
Ref.~\cite{RV-21} and references therein.  As shown for various
systems and modelizations of the dissipative interactions within the
Lindblad framework, the interaction with the environment makes the
quantum critical behavior of a closed system generally
unstable~\cite{YMZ-14,NRV-19,RV-20,RV-21}, similarly to a finite
temperature. Indeed, in this framework the dissipative interactions
are relevant perturbations, which move the open system away from the
quantum critical behavior of the isolated system.  An alternative
mechanism leading to dissipation is provided by the coupling of a
many-body system with an infinite set of harmonic oscillators, see,
e.g., Refs.~\cite{CL-83,LCD-87,WTS-04,SWT-04,WVTC-05}. Also this type
of dissipative interactions is a relevant perturbation of the critical
behavior of isolated systems, and may lead to other types of
criticality driven by dissipation.  We note that our modelling, in
which the global system is isolated and evolves unitarily, is
physically different from the one used in the Lindblad framework or
when an infinite set of oscillators is coupled with the system.

The predicted FSS behaviors put forward in this paper have been
clearly observed in numerical simulations of stacked Ising chains for
a relatively small number of coupled spins. For example, the scaling
behavior of the decoherence factor has been observed in systems with a
few tens of qubits.  This suggests the possibility of devising
realistic experiments with quantum simulators to address the phenomena
discussed here, by means of various platforms, see, e.g.,
Refs.~\cite{Islam-etal-11, Debnath-etal-16, Simon-etal-11,
  Labuhn-etal-16, Salathe-etal-15, Cervera-18, Keesling-19}.

We finally remark that the results obtained in weakly coupled
$d$-dimensional quantum Ising systems also apply to the corresponding
classical systems, i.e., to coupled $D$-dimensional classical Ising
systems with $D=d+1$. For instance, we may consider two coupled
$D$-dimensional lattice Ising systems ${\cal S}$ and ${\cal E}$,
defined by the partition function
\begin{eqnarray}
&&  Z = \sum_{{\bm s_{\bm x}}} \exp(-H_{cl}/T) \,,\label{clpartfun}\\
&& H_{cl} = H_{\cal S}(J)  +  H_{\cal E}(J_e) + H_{\cal SE}(\kappa) \,,\nonumber
\end{eqnarray}
where 
\begin{eqnarray}
&& H_{\cal S}(J) = - J \sum_{\langle {\bm x}{\bm y}\rangle} s_{{\bm
      x}} \, s_{{\bm y}}\,, \quad H_{\cal E}(J_e) = - J_e\sum_{\langle
    {\bm x}{\bm y}\rangle} w_{{\bm x}} \, w_{{\bm y}}\,,
  \nonumber\\ &&H_{\cal SE}(\kappa) = -\kappa \, \sum_{\bm x} s_{{\bm
      x}} \, w_{{\bm x}}\,. \label{classham}
\end{eqnarray}
Here $s_{{\bm x}}=\pm 1$ and $w_{{\bm x}}=\pm 1$ are classical spin
variables associated with the sites ${\bm x}$ of a $D$-dimensional
cubic lattice, ${\langle {\bm x} {\bm y} \rangle}$ indicates
nearest-neighbour sites.  Using the quantum-to-classical
mapping~\cite{Sachdev-book,RV-21}, the quantum critical behavior of
$d$-dimensional stacked quantum systems coincides with that of
$D$-dimensional stacked classical systems with $D=d+1$.  Then, using
RG arguments analogous to those reported for the stacked quantum Ising
systems (\ref{twosys}) in Sec.~\ref{smallqsca} and \ref{dynsca} (in
the dynamic case, one should also specify a particular dynamics, for
instance, the purely relaxational dynamics~\cite{HH-77}), one can
straightforwardly derive similar FSS relations for the subsystem
${\cal S}$ in the background of the {\em environment} ${\cal E}$, when
${\cal S}$ is close to criticality and the coupling $\kappa$ is
sufficiently small.  These scaling behaviors crucially depend on the
effective phase of the environment ${\cal E}$ controlled by its
parameter $J_e$. Although the scaling behavior is expected to be
analogous, classical and quantum systems are expected to show
significant quantitative differences. For instance, for
$\kappa\to\infty$ the classical model turns out to be equivalent to a
single classical Ising model (in the limit $\kappa\to\infty$ only
configurations satisfying $s_{{\bm x}}=w_{{\bm x}}$ on all sites are
allowed), with an ordered and a disordered phase separated by a
standard Ising transition. In the quantum case, for large $\kappa$ the
system can also be modelled by a single Ising chain, see
App.~\ref{AppB}.  However, the width of the paramagnetic phase shrinks
as $\kappa$ increases, so that the two chains are always ordered in
the limit $\kappa\to\infty$.

\appendix
\section{Landau-Ginzburg-Wilson mean-field analysis} \label{AppA}

In this Appendix we discuss the model using the standard
Landau-Ginzburg-Wilson approach. We consider a classical model in
$D=d+1$ dimensions with two scalar fields $\phi_1$ and $\phi_2$, with
interaction potential
\begin{equation}
V = \int d^{D} x\, \left[V_2(\phi_1,\phi_2) + 
   V_4(\phi_1,\phi_2) \right]\,,
\end{equation}
where 
\begin{equation} 
V_2(\phi_1,\phi_2) = {r_1\over 2} \phi_1^2 + {r_2\over 2} \phi_2^2 + 
\kappa \, \phi_1 \phi_2\,,
\end{equation}
and $ V_4(\phi_1,\phi_2)$ is the quartic potential.  Cubic terms
do not enter because of the symmetry under simultaneous changes of the
sign of the two fields, $\phi_{1,2} \to - \phi_{1,2}$.
Close to the critical point,
the parameters $r_1$ and $r_2$ correspond to
$g_e-g_{\cal I}$ and $g-g_{\cal I}$, respectively.

In the mean-field approach, the kinetic term is neglected and the
transition lines are determined from the analysis of the quadratic
terms. To clarify the behavior, we first perform a unitary
transformation of the fields that diagonalizes the quadratic part. If
$\psi_1$ and $\psi_2$ are the new fields we obtain
\begin{eqnarray}
V_2 &=& {1\over 4} (r_1 + r_2 + \delta) \psi^2_1 + 
      {1\over 4} (r_1 + r_2 - \delta) \psi^2_2\,,
\nonumber \\
\delta &=&  \sqrt{(r_1-r_2)^2 + 4 \kappa^2}\,. 
\end{eqnarray}
Therefore phase transitions occur for 
\begin{equation}
   r_1 + r_2 = \pm \delta \quad \Rightarrow \quad r_1 r_2 = \kappa^2\,.
\end{equation}
This relation implies that, if one of the two systems is at
criticality, for instance $r_1 = 0$, in the $(r_2,\kappa)$ plane there is
a single transition point at $r_2 = \kappa = 0$. Otherwise, we observe
transition lines along which $r_1$ and $r_2$ have the same sign. In
our quantum model, this implies that transitions lie in the region $g>
g_{\cal I}$, $g_e> g_{\cal I}$, for $\kappa \not = 0$. The mean-field
argument also allows transition with $g< g_{\cal I}$, $g_e< g_{\cal
  I}$. However, in this case, the two systems are already ordered, and
thus the additions of the coupling between the two systems would
simply increase the order, without giving rise to critical
transitions.

To understand the nature of the transitions, we should analyze the
quartic potential. We do this analysis for the simple case, in which
the model is also symmetric under the exchange of the two Ising fields (in
the quantum model it corresponds to $g_e = g$). In the presence of
this additional symmetry, the quartic potential is given by
\begin{equation}
V_4 = a_1 (\phi^4_1 + \phi_2^4) + a_2 (\phi^2_1 + \phi_2^2) \phi_1 \phi_2 + 
      a_3 \phi_1^2 \phi_2^2\,.
\end{equation}
In terms of the fields $\psi_i$ it becomes
\begin{eqnarray}
V_4 &=& {1\over 4} (2 a_1 + a_3 + a_2 s_\kappa) \psi_1^4 + 
      {1\over 4} (2 a_1 + a_3 - a_2 s_\kappa) \psi_1^4 
\nonumber \\ 
    && +   {1\over 2} (6 a_1 - a_3) \psi_1^2 \psi_2^2\,,
\end{eqnarray}
where $s_\kappa = \kappa/|\kappa|$.  The two ${\mathbb Z}_2$
symmetries of the original model imply that the system is invariant
under independent changes of the sign of $\psi_1$ and $\psi_2$, which
explains why there are no odd powers of $\psi_i$ in the quartic
potential. Thus, in the symmetric case, the generic model corresponds
to two Ising systems with an energy-energy coupling. For systems in
$D=3$ dimensions there is the possibility of a symmetry enlargement at
the multicritical point---the ${\mathbb Z}_2\otimes {\mathbb Z}_2$
invariance enlarges to O(2)~\cite{FN-74,NKF-74,CPV-03,BPV-22}. In
$D=2$ dimension the energy-energy interaction is marginal at the O(2)
fixed point and thus a more complex behavior can be obtained. As we
have discussed in the text, the latter possibility can be realized by
adding a coupling between the transverse spins, such as in
Eq.~(\ref{altint}).  The model we consider has a decoupled
multicritical point, i.e., it corresponds to $a_3 = 6 a_1$. Thus, the
effective model is simply the sum of two noninteracting Ising
systems. Thus, at fixed $\kappa\not=0$, the transition is obtained by
tuning $g=g_e$ at a critical point $g_c(\kappa)$. The quantity
$g-g_c(\kappa)$ is a thermal Ising scaling field. Nothing would change
for $g\not= g_e$. Odd terms $\psi_1 \psi_2^3$ and $\psi_3 \psi_1$,
would now be present. However, for $\kappa\not= 0$, only one field is
critical.  The noncritical field should be integrated out and it would
give rise to even contributions in the critical field. Thus, we
predict all deviations from the critical point to represent Ising
thermal scaling fields.

\section{Large-$\kappa$ and large-$g_e$ transitions} \label{AppB}

In this Appendix, we discuss the behavior of the system for large values of 
$\kappa$ and for large values of $g_e$.

Let us first consider the limit $\kappa\to\infty$ at fixed $g$ and $g_e$. 
If the system is 
in the global ground state,
the coupling term $H_{\cal SE}$ forces the ${\cal S}$ and ${\cal E}$ 
spins on the same site to
be aligned. Therefore, we can simply consider the problem
in the reduced Hilbert space obtained by only considering the states
$|++\rangle_1$ and $|--\rangle_1$ on each site. 
Here $|++\rangle_1$ and $|--\rangle_1$  are the single-site eigenvectors of
$\sigma^{(1)}$ and $\tau^{(1)}$: $\sigma^{(1)} |++\rangle_1 = +|++\rangle_1$,
$\tau^{(1)} |++\rangle_1 = +|++\rangle_1$, and 
$\sigma^{(1)} |--\rangle_1 = -|--\rangle_1$,
$\tau^{(1)} |--\rangle_1 = -|--\rangle_1$.
The computation of the 
ground state of the total Hamiltonian in this reduced space is trivial. 
The ground state is independent of $g_e$ and $g$ and doubly degenerate:
a basis is provided by the two fully ordered configurations. 

In this derivation we have assumed that $g$ and $g_e$ are fixed as $\kappa$
increases, i.e. we are in the limit $g,g_e\ll \kappa$. However, it is equally 
possible that $\kappa$ is large, but still smaller than $g$, i.e., 
the couplings satisfy $g_e \ll \kappa \ll g$. In this case, the relevant states 
will be eigenstates of $\sigma^{(3)}_x$ on each site. The coupling Hamiltonian
$\kappa$ does not play any role---the two systems decouple---and one has 
an effectively disordered 
subsystem $\cal S$. Thus, for $\kappa$ and $g$ large, depending on their
relative size, $\cal S$ can be either ordered or disordered. We thus 
expect a transition when $\kappa$ and $g$ are both large, but of the same
order.

For convenience, we will adopt below the following notation for the 
single-site states. We indicate with $|++\rangle$, $|+-\rangle$,$|-+\rangle$,
$|--\rangle$ the single-site eigenvectors of $\sigma^{(3)}$ and $\tau^{(3)}$,
respectively.  For instance, $|+-\rangle$ satisfies 
$\sigma^{(3)} |+-\rangle = +|+-\rangle$ and 
$\tau^{(3)} |+-\rangle = -|+-\rangle$. 
If both $\kappa$ and $g$ 
are large, the relevant Hilbert space can be obtained by diagonalizing 
the single-site Hamiltonian, i.e., by considering 
\begin{equation}
H_{ss} = - g \sigma^{(3)} - g_e \tau^{(3)} - \kappa \sigma^{(1)}\tau^{(1)}.
\end{equation}
For $\kappa,g\gg g_e\sim 1$ the relevant eigenstates are 
\begin{eqnarray}
\psi_0 &=& {A\over \sqrt{1 + A^2}} |++\rangle + 
     {1\over \sqrt{1 + A^2}} |--\rangle , \nonumber \\
\psi_1 &=& {A\over \sqrt{1 + A^2}} |+-\rangle + 
     {1\over \sqrt{1 + A^2}} |-+\rangle ,
\end{eqnarray}
where, neglecting corrections of order $g_e/g$, we have 
\begin{equation}
A = {g\over \kappa} + \sqrt{1 + {g^2\over \kappa^2}}.
\end{equation}
The gap is 
\begin{equation}
    \Delta = E(\psi_1) - E(\psi_0) = {2 g\over \sqrt{g^2 + \kappa^2}} g_e,
\end{equation}
where we have only kept the leading term in $g_e$. 
The other two states have energy differences of order $g,\kappa$, with respect
to the ground state. The previous expression shows the presence of two
different regimes, in agreement with the previous discussion. 
If $\kappa/g \to \infty$, $\Delta$ goes to zero, so that the two states
become degenerate. Moreover, since $A\to 1$, 
the combinations $\psi_0\pm \psi_1$
correspond to the aligned states we have discussed above. 
If instead $g/\kappa \to \infty$, the system is gapped and the ground state  
corresponds to a paramagnetic $\cal S$. 

To discuss the generic case, let us note that, for $g,\kappa\gg g_e$,
 on each site the relevant Hilbert space consists of the 
two states $\psi_0$ and $\psi_1$. If we associate the vector $(1,0)$ to
$\psi_0$ and the vector $(0,1)$ to $\psi_1$, we can write the local 
Hamiltonian in this restricted Hilbert space as 
\begin{equation}
H_{\rm loc} = - {\Delta\over 2} \lambda^{(3)} + E_m \qquad 
\lambda^{(3)} = \begin{pmatrix} 1 & 0 \\ 0 & -1 \end{pmatrix}
\end{equation}
where $\lambda^{(3)}$ is a Pauli matrix and $E_m = [E(\psi_0) +
  E(\psi_1)]/2$.  The hopping part of the full Hamiltonian can be
similarly written as an effective hopping term involving
$\lambda^{(1)}_{\bm x} \lambda^{(1)}_{\bm y}$ on neighbouring sites.
Thus, apart from an additive constant, we end up with an effective
Ising chain with Hamiltonian
\begin{equation}
H = - J_{\rm eff} \sum_{(\bm xy)} \lambda^{(1)}_{\bm x} \lambda^{(1)}_{\bm y} 
    - {\Delta\over 2} \sum_{\bm x} \lambda^{(3)}_{\bm x},
\end{equation}
where 
\begin{equation}
J_{\rm eff} = {1 + 6 A^2 + A^4\over (1 + A^2)^2}.
\end{equation}
Therefore, we predict a transition for 
\begin{equation} 
    J_{\rm eff} = {\Delta\over 2}.
\end{equation}
This equation depends only on the ratio $\kappa/g$ and $g_e$. Therefore,
we predict
\begin{equation}
  \kappa_c(g,g_e) = \alpha(g_e) g\,\quad{\rm for}\; \kappa,g\gg g_e\,.
  \label{kgge}
  \end{equation}
Numerically, we find $\alpha(g_e) = 0.931, 1.99, 5.00$ for $g_e =
2,4,10$.  For large values of $g_e$, we have approximately
$\alpha(g_e) \approx g_e/2 (1 - 2/g_e^{-4})$.  For $g_e \to 1$ we have
$\alpha(g_e) \approx \sqrt{2\over3} (g_e-1)^{1/2}$.  As expected, no
solution exists for $g_e < 1$.

We have performed a numerical check of the prediction by computing
$\kappa_c(g)$ for $g_e=2$ and several values of $g$. Data for $g\ge 2$
are well fitted by $\kappa_c(g) \approx a g + b$, with $a = 0.94(1)$
and $b = -0.5(1)$. The estimate of $a$ is in agreement with the
prediction 0.931, obtained above.

The behavior for large values of $g_e$ at fixed $\kappa$ and $g$ can
be discussed analogously. For $g_e\to \infty$, as discussed in
Sec.~\ref{disordtranline}, the subsystems $\cal E$ and $\cal S$ are
decoupled and $g_c(\kappa) = 1$ for all values of $\kappa$. We wish
now to compute the corrections to this result. We proceed as before,
diagonalizing the local single-site Hamiltonian. For $g_e\to \infty$
the relevant eigenvectors are
\begin{eqnarray}
\psi_0 &=& {B_1\over \sqrt{1 + B_1^2}} |++\rangle + 
     {1\over \sqrt{1 + B_1^2}} |--\rangle , \nonumber \\
\psi_1 &=& {B_2\over \sqrt{1 + B_2^2}} |+-\rangle + 
     {1\over \sqrt{1 + B_2^2}} |-+\rangle ,
\end{eqnarray}
where, to leading order in $g_e$, we have 
\begin{equation}
B_1 = {2 g_e\over \kappa}, \qquad B_2 = {\kappa\over 2 g_e}.
\end{equation}
As before, the states $|++\rangle$, $\ldots$, are eigenstates of
$\sigma^{(3)}$ and $\tau^{(3)}$. In the same limit the gap is
\begin{equation}
\Delta \approx 2 g - {g \kappa^2\over g_e^2}.
\end{equation}
To complete the calculation, we should compute the coupling $J_{\rm
  eff}$ that parameterizes the hopping term. We find
\begin{equation}
J_{\rm eff} =
1 + {4 B_1 B_2\over (B_1^2 + 1) (B_2^2 + 1)} \approx 
1 + {\kappa^2\over g_e^2}.
\end{equation}
Requiring $\Delta/2 = J_{\rm eff}$ we obtain 
\begin{equation}
g_c(\kappa) = 1 + {3\over 2 g_e^2} \kappa^2.
\end{equation}
which is valid as long as $\kappa \ll g_e$.

\end{document}